# Accelerated Medicines Development using a Digital Formulator and a Self-Driving Tableting DataFactory

Faisal Abbas[1,^] & Mohammad Salehian[1,^], Peter Hou[1], Jonathan Moores[1], Jonathan Goldie[1], Alexandros Tsioutsios[1], Victor Portela[2], Quentin Boulay[3], Roland Thiolliere[3], Ashley Stark[4], Jean-Jacques Schwartz[4], Jerome Guerin[4], Andrew G. P. Maloney[5], Alexandru A. Moldovan[5], Gavin K. Reynolds[6], Jérôme Mantanus[7], Catriona Clark[1], Paul Chapman[2], Alastair Florence[1], Daniel Markl[1*]

^ Authors contributed equally.
* Corresponding author, Email: daniel.markl@strath.ac.uk

[1] Centre for Continuous Manufacturing and Advanced Crystallisation (CMAC), Strathclyde Institute of Pharmacy and Biomedical Science (SIPBS), University of Strathclyde, Glasgow, G1 1RD, UK
[2] Glasgow School of Art, Glasgow, G3 6RQ, UK
[3] Medelpharm, ZAC des Malettes, 615 Rue du Chat Botté, 01700 Beynost, France
[4] DEC Group, Chemin du Dévent 3, 1024 Ecublens, Switzerland
[5] The Cambridge Crystallographic Data Centre, 12 Union Road, Cambridge, CB2 1EZ, UK
[6] Sustainable Innovation & Transformational Excellence (xSITE), Pharmaceutical Technology & Development, Operations, AstraZeneca UK Limited, Macclesfield, SK10 2NA, UK
[7] UCB S.A, 60 Allée de la Recherche, 1070 Brussels, Belgium

**Abstract**

Pharmaceutical tablet formulation and process development, traditionally a complex and multi-dimensional decision-making process, necessitates extensive experimentation and resources, often resulting in suboptimal solutions. This study presents an integrated platform for tablet formulation and manufacturing, built around a Digital Formulator and a Self-Driving Tableting DataFactory. By combining predictive modelling, optimisation algorithms, and automation, this system offers a material-to-product approach to predict and optimise critical quality attributes for different formulations, linking raw material attributes to key blend and tablet properties, such as flowability, porosity, and tensile strength. The platform leverages the Digital Formulator, an in-silico optimisation framework that employs a hybrid system of models – melding data-driven and mechanistic models – to identify optimal formulation settings for manufacturability. Optimised formulations then proceed through the self-driving Tableting DataFactory, which includes automated powder dosing, tablet compression and performance testing, followed by iterative refinement of process parameters through Bayesian optimisation methods. This approach accelerates the timeline from material characterisation to development of an in-specification tablet within 6 hours, utilising less than 5 grams of API, and manufacturing small batch sizes of up to 1,440 tablets with augmented and mixed reality enabled real-time quality control within 24 hours. Validation across multiple APIs and drug loadings underscores the platform's capacity to reliably meet target quality attributes, positioning it as a transformative solution for accelerated and resource-efficient pharmaceutical development.

# 1  Introduction

A wave of Artificial Intelligence (AI) enabled [1,2] technologies for drug discovery and clinical trials are transforming the way new medicines are discovered and evaluated for efficacy and safety. AI-native drug discovery companies generated an average annual drug pipeline growth rate of around 36% from 2010 to 2021 [3], showing an exponential rise of candidates coming to clinical trials. Most global pharmaceutical companies partner with AI technology providers to accelerate drug discovery and clinical research to exploit emerging digital technologies in drug target identification[4,5], generative molecular design[6], automating discovery[7], clinical study protocol optimisation[8], selection of optimal subpopulations[9], dose optimisation [10], therapeutic drug monitoring and dynamic personalised therapy[11], and reducing adverse drug reaction [12]. It is estimated that these scientific advances can shorten drug development timelines from approximately 12-15 years to 3-4 years [13]. The significant scientific and industrial progress in drug discovery and clinical research repositions the main bottleneck in bringing new medicines to patients efficiently. Specifically they position the development processes for Chemistry, Manufacturing, and Controls (CMC), essential for securing regulatory approval for a new drug, on the critical path for registration of new medicines. CMC includes the development of manufacturing routes and processes, formulation, scale-up and a robust control strategy that reliably provides quality products to patients. This development process involves numerous dependent decisions to transform a new drug candidate into a final product that can be manufactured at scale and meets the target product profile (TPP). From a drug product for first in human (FIH) clinical trials to commercialisation, the financial and time penalties of changing a decision rise exponentially over the course of development using current procedures [14]. CMC development processes must therefore adapt to follow the advances in drug discovery and clinical research and ultimately shorten timelines while ensuring product quality and safety. Digitalisation of CMC processes by the utilisation of digital tools such as Big Data generation, predictive modelling and artificial intelligence (AI), automation and robotics can help speed drug development timelines by reducing the experimental burden and increasing lab efficiency. Roughly two-thirds of medicines are administered orally, and approximately half of these medications are in the form of a tablet [15]. Small molecule drugs continue to represent the majority of the pipeline with 52% drugs (29 out of 55) approved by the U.S. Food and Drug Administration (FDA) in 2023 classified as small molecule[16], in line with the five year average for this modality[17]. Most small molecule active pharmaceutical ingredients (APIs) coming through the pipeline lack desirable raw material characteristics such as good flowability, compressibility, compactability and solubility that are necessary for ease of manufacture and meeting the TPP targets [18,19]. This puts additional pressure on making informed decisions on formulation, manufacturing route and processes.

Process and formulation design follows a cycle of hypothesis generation, experimental design, lab testing, and data interpretation to understand effects of various formulation and process factors and design a consistent high-quality product. Yet, this process is often inefficient, repetitive, and time-consuming, potentially spanning months to years. Recent advancements in predictive modelling have been applied across CMC activities, with a particular focus on direct compression (DC) of pharmaceutical tablets, due to its simplicity, cost-effectiveness, and suitability for heat- and moisture-sensitive APIs [20]. DC eliminates the need for granulation steps, thereby reducing production time and preserving the integrity of sensitive compounds [21]. However, formulators often consider granulation due to DC's high dependency on the material properties of the API and excipients (e.g. flowability, compressibility, and compactability). This challenge can be addressed through an autonomous, resource-efficient approach using industrial digital technologies (IDTs) aligned with the Manufacturing Classification System (MCS) [18,22,23] to proactively scope the likelihood of using DC as a viable processing option for given a formulation to meet the TPP.

Mechanistic models – including first-principle and empirical [24-27] – have been utilised to predict tablet formulation and processing parameters. While these models provide valuable insights into the underlying physical and chemical processes, they require extensive experimental data and in-depth physics-based knowledge of the problem to estimate their parameters accurately, demanding significant effort in preparing and calibrating formulations and tablets with often partial domain knowledge. Data-driven models, leveraging machine learning [28-31], deep learning, and computer vision [32-35], offer alternative approaches by identifying complex patterns within experimental data. However, the data processing and training/testing pipelines often lack generalisability. This highlights the need for a hybrid approach that integrates mechanistic understanding with data-driven techniques, minimising experimental burdens while predicting the drug product properties directly from raw material attributes [36].

Another critical aspect in the development process is the systematic optimisation of decision parameters, including formulation compositions and process configurations, to achieve the desired TPPs. This requires researchers to navigate the complex interplay of variables, making informed decisions that balance quality attributes, regulatory requirements, and manufacturing efficiencies [37,38]. Bayesian optimisation methods powered by Gaussian processes have been tested in pharmaceutical process engineering due to their computational efficiency [39,40], however, they are typically used for process optimisation cases with a limited number of decision parameters and may not effectively handle larger-scale problems with numerous variables and different types of input features [41,42]. Therefore, advancing modelling decision-making strategies by combining current methodologies with more robust approaches such as gradient-based [43,44] and gradient-free [45,46] optimisation methods is crucial for addressing the correlation between different types of decision variables in tablet formulation and manufacturing processes.

Self-driving labs have gained significant attention in recent years for their potential to revolutionise material discovery [47,48]. These labs leverage robotics, automation, machine learning, and AI to accelerate discovery and development processes while reducing human error. The closed-loop workflow in a self-driving lab is generally a dynamic operation that cycles through a design, make, test and analyse (DMTA) workflow [49-53]. Relevant examples include a fully autonomous solid-state powder X-ray diffraction (PXRD) workflow using multipurpose collaborative robots [41], self-driving synthesis [54], discovery of heterogeneous catalysis materials [55], and an AI-Chemist platform capable of autonomously performing chemical research tasks, including literature review, experiment design, execution across 14 workstations, and data analysis using machine learning and Bayesian optimisation [56].

Despite this rise in self-driving laboratories and predictive systems for molecular and material discovery, the development of formulations and process conditions for oral drug products remain a largely manual process requiring a significant time investment of subject matter experts [47,57-59]. The emerging era of drug discovery envisions an automated landscape, where workflows encompassing biological assays, chemical synthesis, and data analysis are seamlessly interconnected through versatile, mobile, and modular hardware [60].

In this article, we present a platform that integrates in-silico optimisation using a hybrid system of models and a self-driving Tableting DataFactory (i.e. a digitally integrated cyber-physical infrastructure for tablet manufacturing and testing that systematically collects, processes, and manages data from diverse sources throughout its lifecycle) for accelerated and material-sparing development of the formulation and process conditions of pharmaceutical tablets for a given API. The platform (Figure 1) realises a digital workflow aligned with Quality by Digital Design (QbDD) principles. It begins with the TPP of a *"New Drug (API) Candidate"* that defines a specific target drug loading and manufacturability criteria, and proceeds through several distinct, yet integrated, computational modelling and digitally automated stages to achieve rapid development and manufacture of tablets:

- **Material Characterisation:** The process starts by characterising the material's fundamental properties, including particle size, shape, true density, and bulk density, combined with computed API properties such as particle informatics descriptors, ensuring that the raw material is well-described before proceeding to the in-silico optimisation.

- **Digital Formulator:** Using a hybrid system of material-to-product models, an in-silico optimiser identifies the best formulation (excipient selection and mass fractions) and initial process conditions that maximises the flowability of the material while meeting manufacturability criteria, specifically porosity and tensile strength.

- **Self-driving Tableting DataFactory:** Once the optimised formulation is established, the Tableting DataFactory employs physics-informed and multi-output Bayesian optimisation to refine process conditions and develop knowledge about the impact of process conditions on quality attributes of the tablets. The development pathway, encompassing material characterization, digital (in silico) formulation optimisation, and Bayesian process optimisation using model driven experiments delivered on the Tableting DataFactory, is completed in within six hours.

- **Manufacturing:** The Tableting DataFactory operates in a manufacturing mode capable of producing 1,440 tablets within 24 hours. While the current setup is not designed for Good Manufacturing Practice (GMP) compliance, it serves as a proof-of-concept for applications such as early-phase clinical trials, dose titration studies, or personalised healthcare, where flexibility and adaptability are critical[61-63]. Real-time process monitoring is facilitated by augmented and mixed reality (AR/MR) visualisation tools, enabling parameter monitoring and adjustment to ensure high-quality output in real-time scenarios [64].

The platform was tested with nine different drug loadings across six APIs, demonstrating its ability to rapidly and efficiently develop formulations and processes for DC of tablets. This approach has the potential to significantly reduce the time from initial material characterisation to clinical supply.

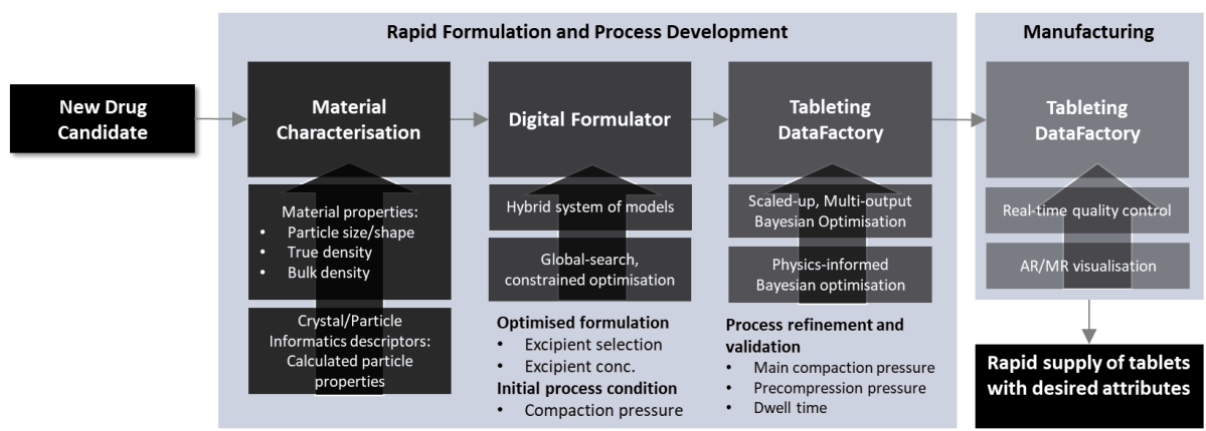

Figure 1: Overview of the platform that integrates material characterisation, in-silico tablet formulation and process development, a self-driving Tableting DataFactory to refine process conditions, and manufacturing of small batches with AR/MR-enabled real-time quality control.

## 2 Digital Formulator

### 2.1 Hybrid System of Models

A dataset comprising 113 tablet formulations (653 data points collected at varying compression pressures; Table S3 in Supporting Information) was employed to develop two material-to-product models predicting porosity and tensile strength of tablets from raw material attributes, formulation descriptors, and process conditions (Figure 2). The system of models is comprised of two connected steps: 1) mixture models [65], 2) process models. The mixture models predict the true density, bulk density, tapped density, particle size distribution, aspect ratio distribution and flow function coefficient (FFC) of a blend of materials with a given formulation (API, excipients and mass fractions) from the raw material's true density, bulk density, particle size distribution and aspect ratio distribution. The process models utilise the output of the mixture models with additional input features about the API (i.e. the particle informatics descriptors and API concentration) and the compaction pressure.

The predictive performance of three modelling approaches was compared, showing the superior performance of the Deep Neural Networks (DNNs) in predicting the porosity and tensile strength of tablets containing previously unseen APIs. This comparison was initially performed without using the particle informatic descriptors (Figure S1 and Table S5 in Supporting Information) followed by using all parameters in Table 3 as input parameters for process models (Figure 3). The distribution of training data overlaid on the DNN predictions indicates that higher standard deviations are associated with regions where fewer training data points were available[66]. Despite these uncertainties, both models demonstrated reasonably low estimated standard deviations in predicted data points for porosity and tensile strength.

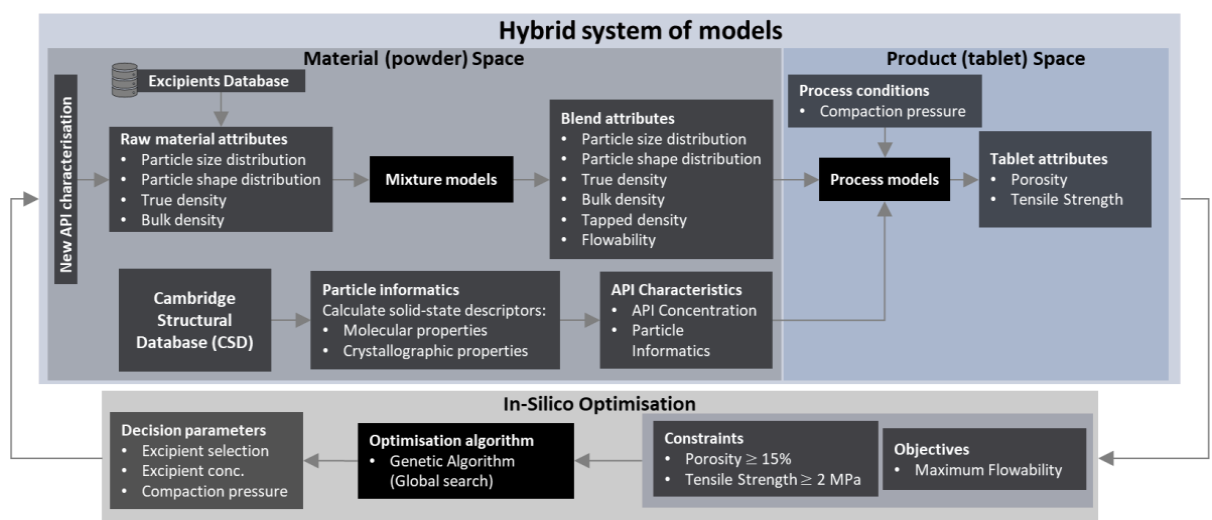

Figure 2: Schematic overview of the Digital Formulator comprised of the hybrid system of models and in-silico optimisation framework.

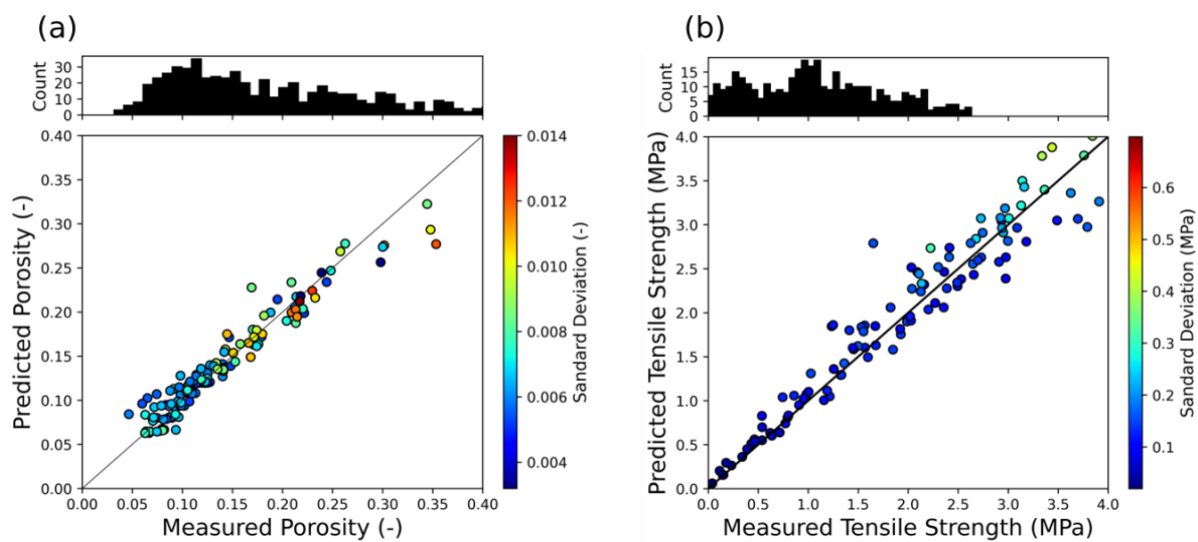

Figure 3: Prediction performance of (a) porosity and (b) tensile strength DNN models using the full set of input parameters, including calculated particle informatics descriptors. The scatter plots compare the predicted versus measured values, while the colour bars represent the standard deviation of predictions. The histograms above each scatter plot show the distribution of training data points for porosity (left) and tensile strength (right).

## 2.2 In-silico Optimisation

The Digital Formulator uses the hybrid system of models in an optimisation framework (Figure 2) to find the optimal set of excipients, their mass fractions (concentrations), and the initial main compression pressure for unseen APIs and a given target drug loading. The objective is to maximise the FFC (i.e., minimise the negative of the objective function) subject to processability constraints. The ensemble learning strategy used to train DNNs allows the prediction of the uncertainty (i.e. standard deviation of predicted values over the ensemble of models) of future predictions offering valuable insights into the trade-off between exploration (i.e., maximising model accuracy) and exploitation (i.e., optimising the target outcome) during the optimisation process. This has been considered by incorporating the uncertainty associated with the predicted porosity and tensile strength in the processability constraints imposed during the optimisation.

For this study, we developed the system using common DC formulations consisting of five components: API, excipient 1, excipient 2, lubricant, and disintegrant, along with their corresponding mass fractions. We selected a sub-set of the excipients (see Table 1 in Supporting Information) for the optimisation which align with standard industry practices to ensure findings are relevant to real-world applications [67,68]. It should be noted that the hybrid system is inherently flexible and can readily accommodate an expanded list of excipients as appropriate, underscoring its adaptability to a wide range of formulation requirements.

To reduce the complexity of the problem, the lubricant and disintegrant were assumed to be constant, magnesium stearate (MgSt; lubricant) and croscarmellose sodium (CCS; disintegrant) with mass fractions of 0.035 (3.5% w/w) and 0.01 (1% w/w), respectively. Table 6 in Supporting Information summarises the decision parameters and their values/ranges. The formulation optimisation cases were repeated for different APIs and target API mass fractions. The APIs investigated include SP (16% w/w), SP (18% w/w), SP (20% w/w), SP (22% w/w), AS (20% w/w), DM (20% w/w), GR (20% w/w), IM (20% w/w), MH (20% w/w), as per listed in Table 1 in Supporting Information. For each API and target drug loading, an optimisation problem was run to optimise the selection and mass fraction of excipients 1 and 2 as well as the initial main compression pressure that maximises the FFC at a constant consolidation pressure (1.6 kPa in this study).

The heatmap (Figure 4) illustrates the optimal concentration profiles of various excipients across the range of formulation optimisation cases. MCC Avicel PH102 emerges as the predominant excipient in several optimal solutions, particularly at higher API loadings of SP, where its mass fraction increases from 49.1% w/w (SP 16% w/w) to 73.5% w/w (SP 22% w/w).

In all cases (except GR 20% w/w where LAC FastFlo 316 has the higher concentration), single or multiple grades of MCC dominate the optimal solutions, demonstrating its significant role in achieving the required tensile strength and porosity within the imposed constraints, attributed to its superior compressibility and binding properties [68,69]. Conversely, other excipients such as LAC Granulac 200M and MAN Pearlitol 200 SD are either minimally utilised or entirely absent from the optimal formulations, indicating their limited contribution to maximising the FFC under the given compressibility constraints. The sporadic, concentration-dependent inclusion of LAC FastFlo 316 and MCC Avicel PH101 suggests that their presence in the formulation is highly sensitive to the specific API and its loading.

There are, however, multiple other factors that needs to be further considered when optimising the formulation. For example, the proportion of MCC can be potentially restricted due its insoluble nature, which can affect the tablet's dissolution rate and API bioavailability [70]. It also has a high moisture content, posing stability issues for moisture-sensitive APIs [71]. Moreover, MCC is strain rate sensitive, leading to inconsistencies in tablet hardness during high-speed manufacturing [72]. Therefore, despite its advantages, it's important to balance MCC's use by considering these factors and potentially exploring alternative excipients to address these challenges. Another parameter is the impact of lubricant and disintegrant on the manufacturability and performance (e.g. disintegration, dissolution) criteria, which remains a promising topic to be studied in a future work.

The comparison between predicted and measured FFC values across different API loadings further validates the optimisation process (Figure S2 in Supporting Information). The predicted FFC values generally trend higher than the measured values, particularly for APIs such as AS (20% w/w) and GR (20% w/w), where the predictions overestimate the FFC. This discrepancy may initially be attributed to the conservative nature of the robust optimisation approach, where the imposed constraints and uncertainty factors lead to an over-prediction of the FFC to ensure that the formulation meets the processability criteria under a range of potential conditions. Furthermore, the error bars on the measured FFC values indicate that the discrepancy between measured and predicted values tends to occur when the uncertainty of measurement is higher, suggesting that the measurement of flowability in these cases is more challenging. Despite being lower, the measured FFC values of all optimal formulations, except DM (20% w/w) and GR (20% w/w) cases, successfully meet the required flowability threshold ($FFC > 4$), validated through the preparation and characterisation of the respective blends. This consistency suggests that, while the predictive model may underestimate values on the side of caution (i.e.

predicting poorer flowability indices than the actual one), it effectively guides the formulation process towards producing robust and viable tablets.

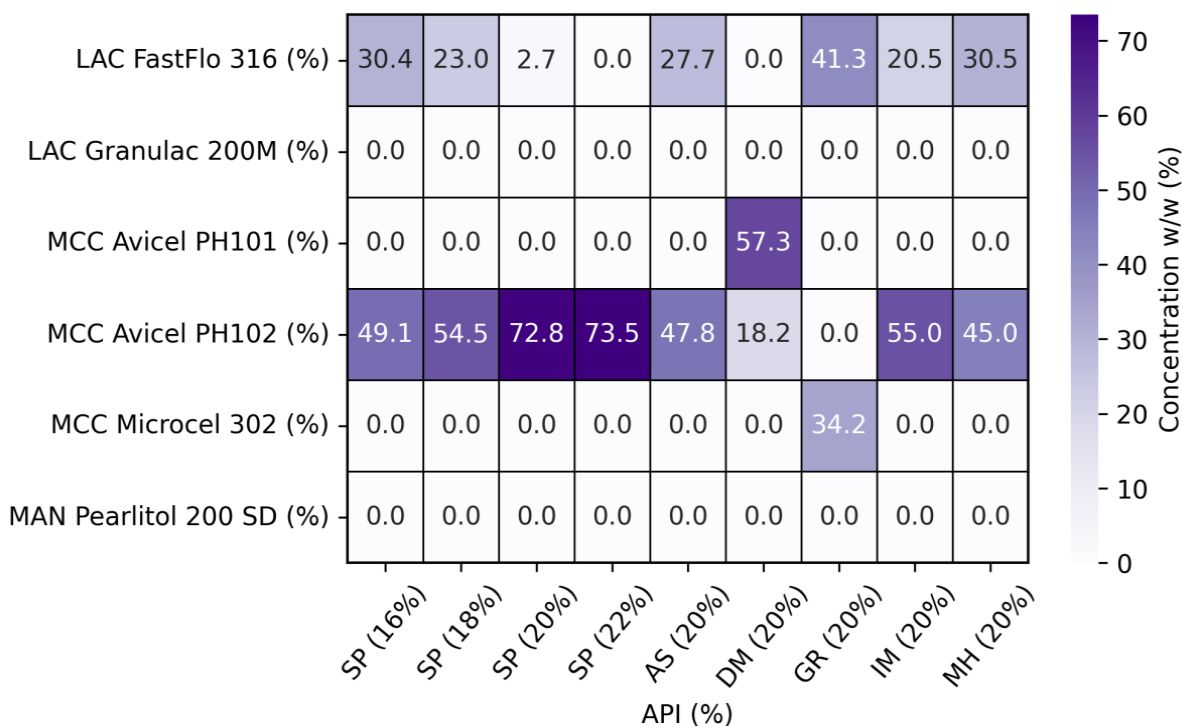

Figure 4: Summary of results from formulation optimisation cases. The $x$-axis represents different case studies with target APIs and their corresponding mass fractions (w/w) as percentages. The $y$-axis lists the excipients used in the formulation optimisation process with their respective optimised concentrations (w/w) shown as percentages. The colour scale indicates the concentration (w/w) of each excipient. The '0.0' indicates that this excipient was not chosen by the Digital Formulator for the given API, objective and constraints.

## 3 Tableting DataFactory

The experimental setup integrates commercially available devices with several customised features as shown in **Figure 5**a. All instruments are digitally interfaced with a LabVIEW-based supervisory control unit (SCU). Robotic arm 1 (R1) is employed for material transport and to physically interconnect all devices. Each iteration of the tablet production begins with the autonomous acquisition of a near-infrared spectroscopic dark scan and a reference scan, using a 99% reflectance disc permanently attached to the R1 gripper as shown in **Figure 5**a. The SCU captures the reference scan once R1 positions the disc on the near-infrared spectrometer (NIRS). Subsequently, R1 proceeds to the dosing unit to acquire the dose for a single tablet. The dosing unit, preloaded with a premixed powder blend, dispenses the precise amount of powder required for one tablet into a customised 3D-printed transportation unit (TU) that is placed on a balance. (Figure S3 in the Supporting Information).

The first decision point (D1) concerns the weight of the obtained powder, measured in real-time (Figure 1C). Any dose deviating by more than ±5% from the target is rejected and subsequently recycled. The powder dose that satisfies the weight thresholds moves on the NIRS stage where the spectrum is acquired by placing the TU on the NIRS. The acquired spectrum is used to monitor the blend homogeneity. The assessment of blend homogeneity was performed qualitatively using Hotelling's $T^2$ analysis, which evaluates the multivariate distance of each spectrum from the principal component analysis (PCA) model centre to identify any deviations between iterations. This approach validates the blend subsamples do not deviate significantly from one another, supporting the assumption that the blend is homogeneous. Currently, powder doses are neither diverted to waste nor reused in cases of blend inhomogeneity. However, any produced tablets that do not meet given content uniformity criteria can be identified and discarded if necessary. The absence of frequent outliers or variations between subsamples further indicated a consistent distribution of components within the blend (Supporting Information 3.7.2). R1 then transports the TU to the compaction simulator, depositing the powder into a 9 mm die. Tablets formed in the compaction simulator are conveyed to an automated tablet tester via a customised chute. The automated tablet tester conducts destructive testing on selected tablets, measuring weight, thickness, diameter, and breaking force to determine tablet porosity and tensile strength. The number of tablets undergoing destructive testing is set in the experimental protocol by the researcher. The remaining tablets undergo non-destructive testing, which measures all parameters except breaking force. A customised tablet separator (TS) then segregates the damaged from the undamaged tablets (Fig S6 in the Supporting Information).

The second decision point (D2) assesses whether tablets should be discarded if they fail to meet quality standards such as tablet weight, porosity and tensile strength beyond acceptable variations (± 5%). Undamaged tablets are collected by the robotic arm 2 (R2) using customised gripper fingers and stored in designated containers. Finally, to prevent cross-contamination with other blends, a customised 3D-printed cleaning unit (CU) is used to thoroughly clean the tube and casing of the TU (Figure S7 in Supporting Information).

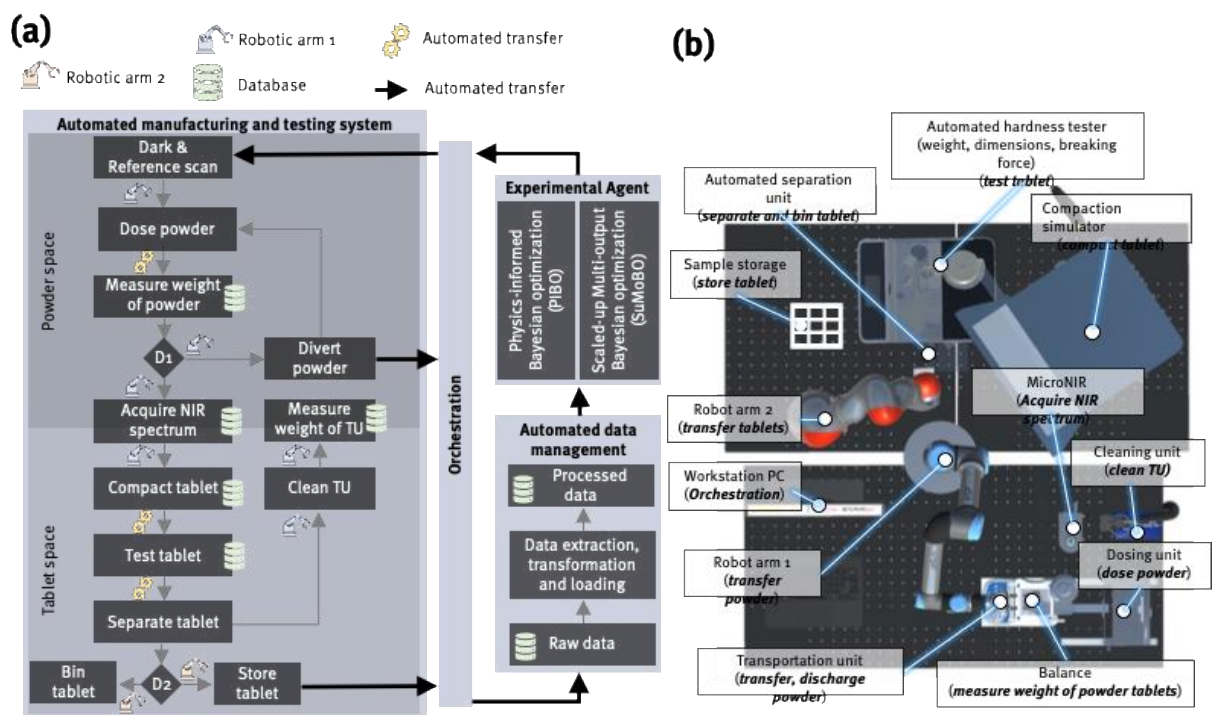

Figure 5: Overview of the Tableting DataFactory. a) The workflow detailing the operation of the automated system alongside the automated data management and experimental agent. D1 is the first decision point to either discard the powder dose or take it forward based on weight of the dose. D2 is the second decision point to either discard the damaged or unacceptable tablets or move them to storage. B) Top view of the setup indicating the location of each instrument on the table. Video 1 in Supplementary Information demonstrates the operation of the Tableting DataFactory.

### 3.1 Benchmarking and Validation

Benchmarking and validation of the system focused on assessing 1) the powder and tablet weight, and any related powder loss caused by the powder handling and transportation, and 2) the consistency of repeated experiments.

The primary objective in the design of the TU was to discharge the powder precisely while keeping the powder loss in a consistent range across different formulations. Therefore, powder loss between the powder obtained, measured by the balance, and the final tablet weight was assessed across ten different blends (B10 – B19) and three different target tablet weights (200, 300 and 400 mg) (Section 3.7 in Supporting Information). The total powder loss during transportation, including losses from powder dosing and spillage when opening the TU gate in tablet press, consistently stays within 10 - 22 mg across different formulations and dose weights (Figs. S9-S12 in Supporting Information). Material is primarily lost in bulk rather than selectively, the ratio of API to excipients remained stable throughout the process (Section 3.7.1 in Supporting Information). Moreover, the relative standard deviation in powder obtained and the tablet weight remains below 3%. In the self-driving tableting mode, the mean deviation from target tablet weight and the porosity for the blends B1-B9 was kept below 4% and 1%, respectively, demonstrating an acceptable consistency across these parameters.

## 4 Experimental Agents for Self-driving Tableting DataFactory

The optimum formulations from the nine case studies (Table 2 in Supporting Information) were used in two real-time optimisation frameworks realising a self-driving Tableting DataFactory (Figure 6):

- Physics-informed Bayesian optimisation (PIBO) framework: PIBO optimises the main compression pressure to meet the target porosity and tensile strength while taking the existing physics-based (PB) correlations between the input and objectives into account, leading to faster convergence and reduced number of experiments required to generate the compressibility and compactability profiles.
- Multi-output Bayesian optimisation (MOBO) framework: MOBO is set up in exploration mode to develop a model that connects multiple intercorrelated process parameters (pre-compression pressure, main compression pressure, and dwell time) and objectives (tablet porosity, tensile strength, and elastic recovery).

Both optimisation frameworks were digitally integrated with the Tableting DataFactory through a local call mechanism between LabVIEW and Python scripts.

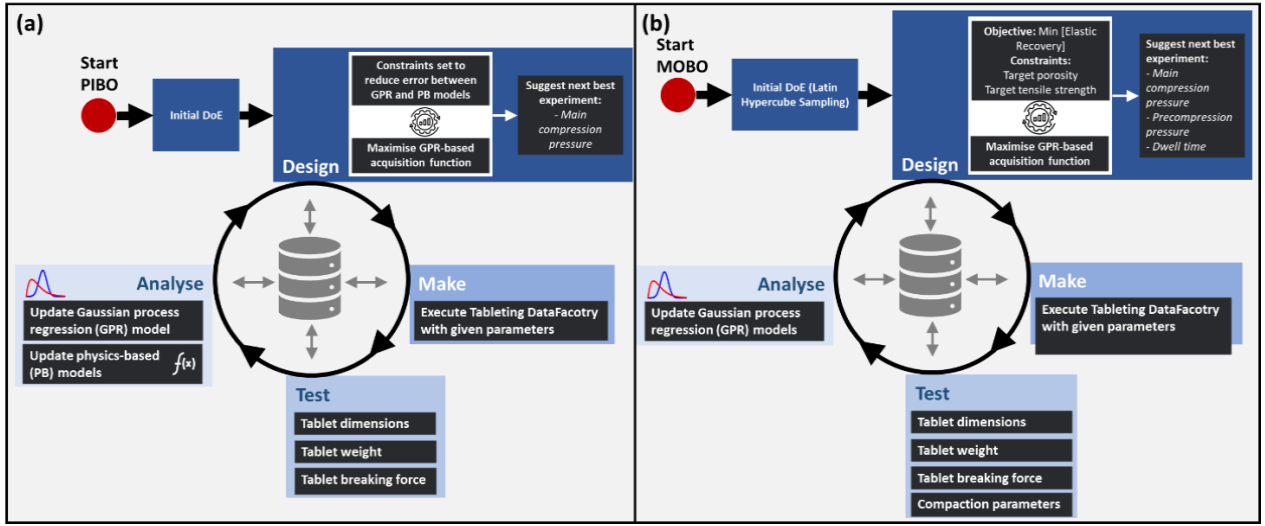

Figure 6: Schematic representation of the two closed-loop DMTA process optimisation workflows for the self-driving Tableting DataFactory. (a) PIBO workflow based on a Gaussian Process Regression (GPR) model and Physics-based (PB) models. To incorporate the compaction behaviour captured by empirical models into the BO process, constraints are set to reduce the error between the two (GPR and PB) models, with the system iteratively suggesting the next experiment and registering new data points, followed by model updates. (b) MOBO workflow with an additional focus on minimising elastic recovery. The optimisation process begins with an initial Design of Experiments (DoE), and subsequent experiments are suggested to develop predictive models for the desired quality attributes. New data points are continuously registered, and the GPR models are updated accordingly. MOBO is terminated when a user-defined number of iterations is reached.

## 4.1 Physics-informed Bayesian Optimisation (PIBO)

The PIBO framework aims to optimise the main compression pressure to achieve the target porosity and tensile strength while satisfying the underlying physics-based (empirical) compressibility and compactability models that provide prior information about the relationship between main compression pressure, porosity, and tensile strength. The PIBO framework was tested with the Tableting DataFactory across the nine formulations identified by the Digital Formulator (Figure 4 in Section 3 and Table 2 in Supporting Information) to achieve target porosity of 0.15, i.e. minimising $Err(\varepsilon_o, \varepsilon_T)$, which is the error between $\varepsilon_o$, the observed porosity, and $\varepsilon_T$, the target porosity by optimising the main compression pressure (see section 7.3.2 for detailed mathematical formulation of the problem). Note that the tensile strength is subsequently assessed as it is a function of porosity and breaking force. Figure 7 shows the variation of tuning parameters of physics-based models during the optimisation process, where all four tuning parameters converge to a plateau after a few iterations, showing the successful performance of the PIBO in general. Moreover, the termination criteria have been activated during different iterations based on the speed of convergence in each case study. The termination of optimisation is followed by a validation experiment at the target porosity, where a compression pressure is suggested based on the calculated tuning parameters and physics-based models.

The measured porosity is compared with the predicted target to evaluate the accuracy of calibrated models. The hybrid system of models, on average, overestimated the porosity by 0.26% and underestimated the tensile strength by 0.63 MPa (see Table 1 for the case-by-case absolute error). The comparison between the initial prediction by the system of models and calibrated models after PIBO demonstrates that the initial predictions (Figure 8 and 9 in Supporting Information), made with historical data[73,74], are reasonably close to the final calibrated models. Both process models capture the overall trends of porosity (i.e. exponentially decreasing as the main compression pressure increases, following the Kawakita model in Eq. 3) and tensile strength (i.e. exponentially decreasing as porosity increases, following Ryshkewitch-Duckworth model in Eq. 4) of optimised formulations even before performing any experiments. While the PIBO calibration process further refines the predictions, the adjustments required for the initially predicted profiles by the system of models require only 6 experiments for the case studies investigated in this work, underscoring the reliability of the DNN-based process models in making a close-to-optimum first-time prediction of tablet attributes from raw material properties. The general underestimation of the initial predictions is due to the conservative definition of the objective function to justify the optimisation in scenarios where extensive experimentation may not be feasible and a right-first-time formulation optimisation is required.

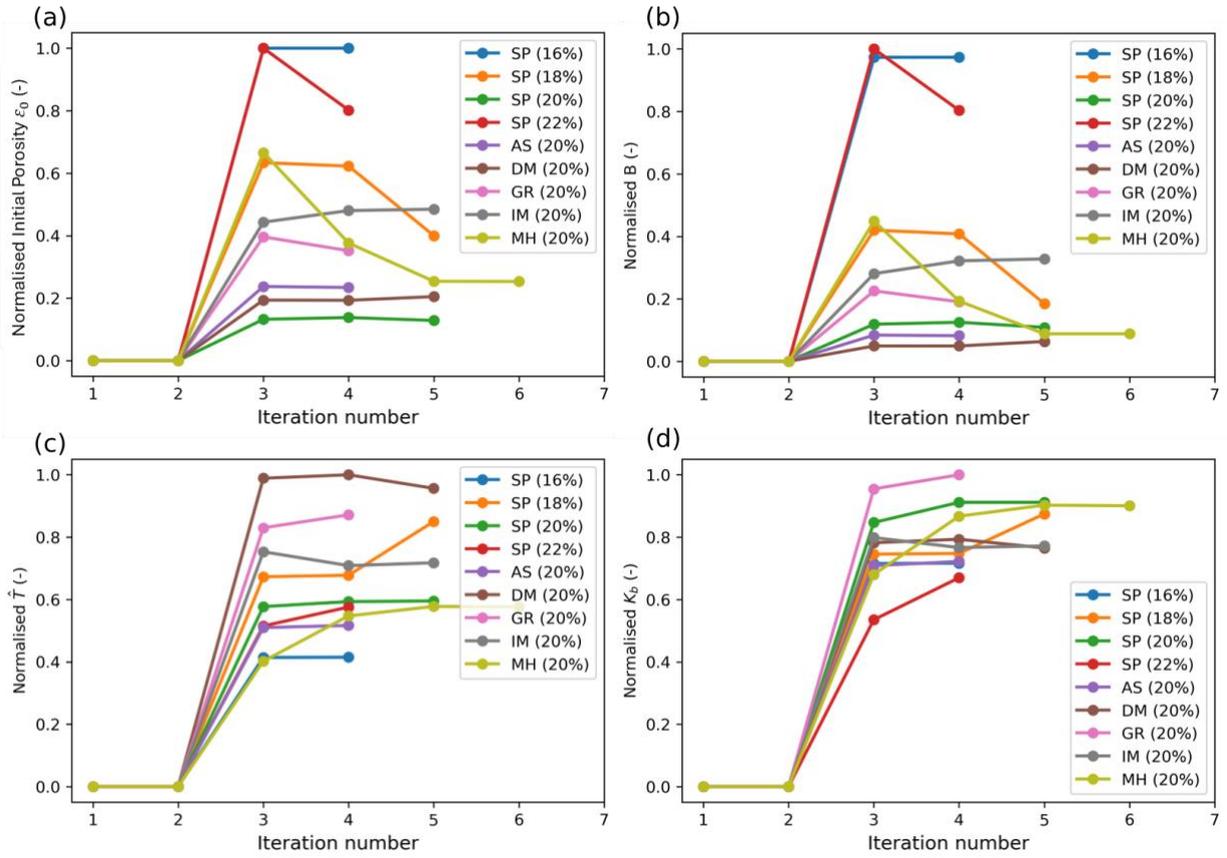

Figure 7: Variation of normalised values of tuning parameters of (a: $\varepsilon_0$, b: $B$) Kawakita and (c: $\hat{T}$, d: $k_b$) Ryshkewitch-Duckworth models during the PIBO cases for different optimised formulations. Each tuning parameter is normalised to the range of 0 and 1 using the minimum and maximum values observed across all iterations.

Table 1: Prediction Error 1: The absolute error between initial model-based prediction (using digital formulator) and experimental result at optimal compression pressure. Prediction Error 2: The absolute error between post-calibrated gaussian process regression model in the PIBO and the experimental result at the optimal compression pressure (validation point). Negative values indicate underestimation, while positive values represent overestimation. Colour intensity indicates deviation from zero: white for values near zero and progressively deeper colours for values further from zero. The red and orange colour scales are independently scaled for porosity and tensile strength values, respectively.

| Case study | Prediction Error 1 (In-Silico vs. validation) | | Prediction Error 2 (PIBO vs. Validation) | |
|---|---|---|---|---|
| | Porosity (-) | Tensile strength (MPa) | Porosity (-) | Tensile strength (MPa) |
| SP (16%) | -0.027 | -0.4 | -0.004 | -0.001 |
| SP (18%) | -0.031 | -2.1 | -0.005 | 0.002 |
| SP (20%) | -0.02 | -0.7 | 0.009 | 0.43 |
| SP (22%) | -0.022 | -2.1 | 0.003 | -0.002 |
| AS (20%) | 0.032 | 0.2 | -0.006 | -0.13 |
| DM (20%) | 0.058 | -0.8 | -0.009 | 0 |
| GR (20%) | 0.005 | -0.1 | 0 | -0.002 |
| IM (20%) | 0.026 | -0.1 | -0.001 | 0 |
| MH (20%) | 0.002 | 0.4 | 0 | 0 |

## 4.2 Multi-output Bayesian Optimisation (MOBO) for Rapid Scale-up Assessment

During scale-up, process parameters can behave differently due to increased speeds and forces, requiring careful consideration of intercorrelation between multiple process parameters [75]. A scale-up, multi-output Bayesian framework was designed to minimise elastic recovery while achieving target porosity and tensile strength by optimising key process parameters such as main compression pressure, precompression pressure, and dwell time. Minimising elastic recovery is critical, as it is closely associated with tablet defects such as lamination, capping, and air entrapment, particularly when scaling up from compaction simulators to rotary tablet presses [76,77].

The proposed MOBO was tested on two formulations, SP (20%) and AS (20%), which were identified using the Digital Formulator. The MOBO was set for exploration, allowing for the collection of sufficient data to effectively train the GP models across a large parameter space (Figure 10 in the Supporting Information). Three individually trained GP models can then be used to predict elastic recovery, tensile strength, and porosity across varying precompression, main compression pressures, and dwell times. This predictive knowledge space is refined to identify a manufacturability region, where tensile strength exceeds 2 MPa and porosity remains above 0.15 (Eq. 8 and 9), as shown in Figure 8. Within this region, the optimal main compression and precompression pressures can be determined to minimise elastic recovery while meeting porosity and tensile strength constraints.

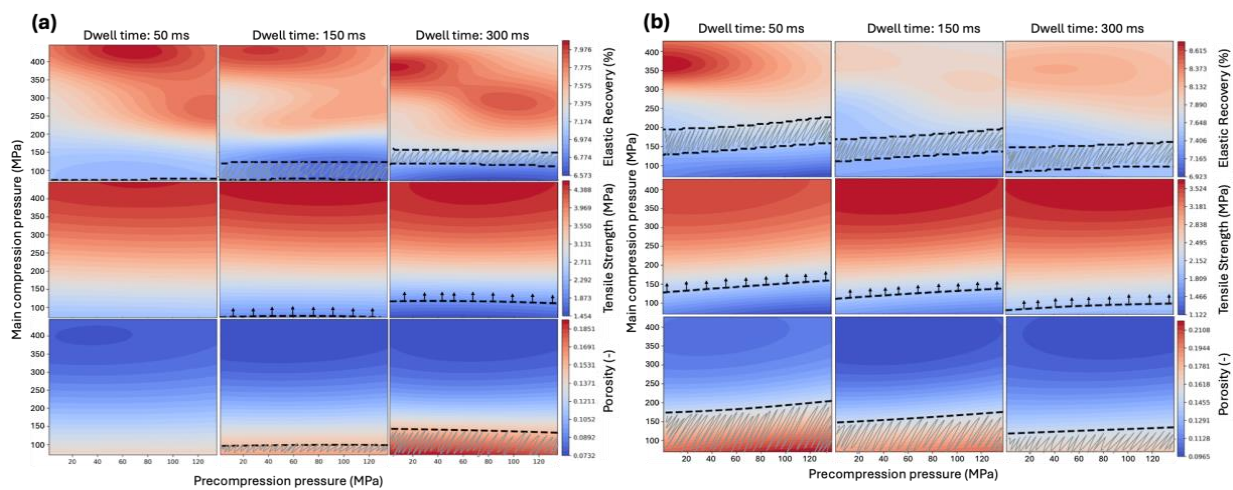

Figure 8: Heatmaps generated by self-driving Tableting DataFactory using MOBO approach. The heatmaps show elastic recovery (top row), tensile strength (middle row), and porosity (bottom row) based on the variation of precompression and main compression pressure at three different dwell times (50, 150, 300 ms) for (a) SP (20%) and (b) AS (20%) case study. For each case study, in the bottom row, the regions filled within the dashed lines correspond to the porosity ≥ 0.15. In the middle row, the arrows in the middle-row figures indicate regions (i.e. above the dashed line) where specific criteria are met, such as the area at the top satisfying tensile strength ≥ 2 MPa. In the top row, the region filled within the dashed lines corresponds to the area that both porosity and tensile strength constrains are satisfied.

## 5  Manufacturing with Extended Reality (XR)-enabled Process Monitoring

The Tableting DataFactory can be operated in a manufacturing mode to deliver a consistent batch of tablets that meet the quality standards. Considering the current speed of the system, it can deliver 1,440 tablets within 24 hours with real-time monitoring of each tablet produced. The system generates multi-dimensional data for every single tablet necessitating novel approaches to support human-driven, data-centric quality monitoring and decision-making. This was achieved by integrating XR technology, specifically augmented reality (AR) and mixed reality (MR) into the Tableting DataFactory to connect physical assets, data and the researcher in an intuitive, accessible and effective manner. AR is a technology that overlays digital content onto the real world, enhancing the user's perception of their environment. MR is the blend of physical and digital worlds where virtual objects interact with real-world elements in real time. However, MR is an umbrella term encompassing AR and MR covering all immersive technologies that merge digital and physical experiences. Both AR and MR platforms visualise a dashboard with the key quality parameters, specifically tablet weight, porosity, and tensile strength in real-time, whilst also streaming real-time data of individual instruments (Figure 9).

AR is utilised in the lab to overlay real-time data directly onto individual instruments. This allows researchers and operators to access crucial information, such as performance metrics, operational statuses, and diagnostic data, without the need to refer to external displays.

The MR version is specifically designed to allow users to visualise experimental data in real-time outside of the laboratory environment. The MR hologram of the Tableting DataFactory provides an immersive, 3D representation of the whole system in operation. This holographic visualisation allows users to virtually observe the production process, inspect the parameters from individual instruments, and troubleshoot potential issues from a remote location. The integration of MR technology thus extends the accessibility and control over the system, significantly improving collaboration, training methods and oversight in a distributed work environment and provides further opportunities for user training. Future work aims to integrate real-time quality control (QC) using extended reality (XR), leveraging the existing holographic infrastructure.

The AR and MR applications in the Tableting DataFactory are demonstrated in the supplementary Videos 2 and 3, respectively. These examples demonstrate continuous production of 100 tablets. The immersive overlaid diagrams in both AR and MR are designed to highlight the acceptable (green) and unacceptable (red) data points. The acceptable range for each of these parameters is ±5% of defined target. The progress of the overall manufacturing stage is shown by visualising the total tablets produced, stored and analysed.

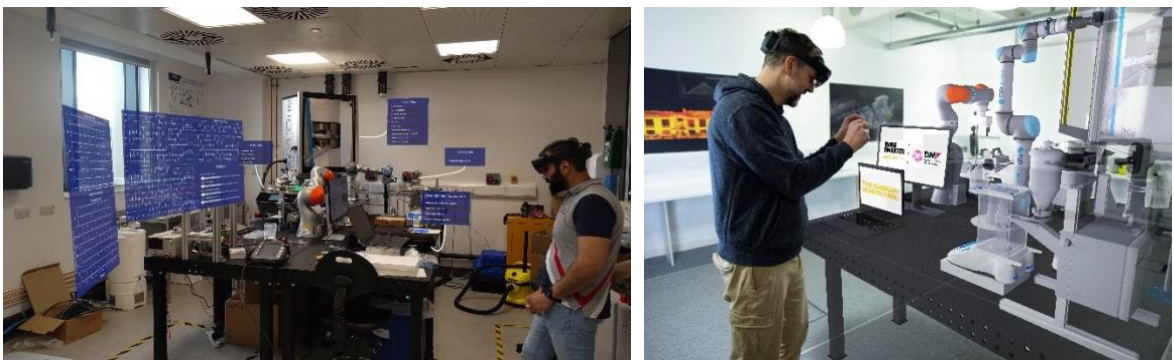

Figure 9: (Left) AR – real-time experimental data overlaid on corresponding equipment during experiment. (Right) MR – Real-time display of holographic equipment for remote laboratory access.

# 6 Conclusions

The Digital Formulator, a model-based optimisation framework, integrates data-driven and mechanistic models to predict and refine optimal formulation settings. The physics-informed Bayesian optimisation framework driving the Tableting DataFactory effectively adjusted process parameters with minimal experimentation, typically fewer than six iterations, while multi-output Bayesian optimisation offered data-driven models to consider interdependencies among key process parameters for scale-up. Further developments in the automated system are envisaged to enable the simultaneous optimisation of formulation and process settings through the adaptive improvement of system of models.

Benchmarking of the Tableting DataFactory demonstrated consistency in both powder and tablet weight, and in minimising powder loss during automated handling, as well as repeatability in manufacturing under fixed process conditions. This automated approach minimises manual intervention, ensuring accuracy, precision, and consistency throughout the production stages. Additionally, augmented and mixed reality integration within the Tableting DataFactory facilitated data-centric quality monitoring and informed decision-making. Table 2 summarises key performance metrics, contrasting the Self-Driving Tabletting DataFactory's operational efficiency with conventional methods, thereby highlighting its potential to reshape pharmaceutical manufacturing.

Combining the Digital Formulator with a Self-driving Tableting DataFactory, delivers optimised tablet formulation and process conditions in under 24 hours while requiring less than 5 grams of API. This study presents a transformative approach to tablet formulation and process development, establishing a resource-efficient and accelerated pathway for formulation and process development of pharmaceutical tablets.

Table 2: Summary of performance metrics of the Digital Formulator with the Tableting DataFactory and comparison with the state of the art (conventional) method, following performance metrics outlined by [78]. The state-of-the-art methods in pharmaceutical industry involve batch processing, manual interventions and do not include automated formulation optimisation as detailed in [79,30,26].

|  | Digital Formulator and Self-driving Tableting DataFactory | State of the Art Method |
| --- | --- | --- |
| **Throughput / productivity gain** | 72 tablets per hour including all the different stages such as powder quality monitoring, tablet manufacturing, testing, storage and data handling. | 10 tablets per hour (manual filling) 20 tablets per hour (shoe feeder) |
| **Material use** | < 5 gr for optimisation of compaction pressure (PIBO) and < 27 gr for the manufacturing of 100 tablets including all the powder losses. | 20 gr for optimisation of compaction pressure (based on minimum blend size) 200 gr for manufacture of 100 tablets (based on blend size) |
| **Human resources** | 6 hours using PIBO with Tableting DataFactory, including: <br>• Material characterisation: 4 hours <br>• Digital Formulator (Predictive system of models and in-silico formulation optimisation): 0.5 hours <br>• Blend preparation: 1 h <br>• Tableting DataFactory with PIBO: 0.5 hour | 14 hours including: <br>• Material characterisation and API compression characterisation [80,26]: 9 hours <br>• Predictive tools to enable product development [79]: 2.5 hours due to manual data handling <br>• Blend preparation: same: 1 hour <br>• Optimise compaction pressure: 1.5 hours |
| **Degree of autonomy** | Human intervention is required only for preparing the blend and filling the hopper of the dosing unit. | Not applicable |
| **Operational lifetime** | Dosing unit's hopper may require refilling every 100 | Not applicable |

|  | tablets as the smaller powder particles may start gathering around the base of the hoper resulting in inaccurate powder dosing. Cleaning of compaction punch may also be required every 1000 tablets. |  |
|---|---|---|
| **Experimental precision** | < 5% of targeted tablet weight. | Not applicable |
| **Accessible parameter space** | Dose weight, pre and main compaction pressure, dwell time, compaction profile, selection of methods for NIR and tablet tester. | Not applicable |
| **Optimisation efficiency** | The PIBO framework converged to the optimal point after 5-6 iterations. The MOBO method required pre-defined number of experiments for the initial DoE (15) and subsequent iterations (25) to develop predictive models for the porosity, tensile strength, and elastic recovery. | Not applicable |

# 7 Methods

## 7.1 Materials

APIs used to produce quaternary and quinary mixture tablets are paracetamol (SP; Standard 6375, Mallinckrodt), acetylsalicylic acid (AS; Molekula), dexamethasone (DM; Powder, Molekula), griseofulvin (GR; Molekula), indomethacin (IM; Molekula), metformin hydrochloride (MH; Molekula). The tablet quaternary and quinary formulations incorporated several filler components, including microcrystalline cellulose with different grades (Avicel PH-101 and Avicel PH-102, DuPont Nutrition) and lactose (FastFlo 316, Foremost Farms USA). Additionally, a specific disintegrant, croscarmellose sodium (CCS) (AcDiSol, FMC International), was added into the formulation. To aid in the compression process, magnesium stearate (Hyqual 5712, Mallinckrodt) was used as a lubricant. Details regarding the characteristics of the excipients and API can be found in Table S1 in Supporting Information. These excipients were chosen to reflect some of the most commonly used materials in DC formulation in industry.

### 7.2 Digital Formulator

#### 7.2.1 Hybrid System of Models

Principal component analysis was performed to reduce the predicted particle size and aspect ratio distributions into two sets of three principal components [81]. The API descriptors include its concentration to explicitly model the effect of drug loading on tablet attributes, and the crystallographic and Particle informatics descriptors represent properties relevant to the processing and mechanical behaviour of the API (Table S4 in Supporting Information)[82]. Table 3 summarises the input features used to train process models. For the tensile strength model, the response variable was log-transformed (i.e., the natural logarithm of tensile strength, $\ln(\sigma_t)$) to incorporate knowledge on the exponential relationship between compression pressure and tensile strength [80].

The formulations feature various APIs (explained in section 7.1), and placebo blends. To evaluate the models' ability to generalise to new APIs, a leave-API-out approach was employed in splitting the dataset into training and test sets. Specifically, 16 formulations containing SP and GR (amounting to 149 data points) were excluded from the training data and reserved as test data. This strategy was designed to assess the models' performance in predicting tablet attributes for APIs not seen during training.

At the initial stage of model development (*Version1*), all input parameters except for informatics descriptors (i.e., parameters in Table S4) were utilised to train data-driven models using three approaches: Deep Neural Network (DNN), Random Forest (RF), and Support Vector Regression (SVR). These approaches were selected to encompass a diverse range of ML-based modelling techniques, where DNN represent deep learning-based approaches capable of capturing complex nonlinear relationships[83], RF is an ensemble-based method known for its robustness and interpretability[84], SVR is a kernel-based approach well-suited for capturing intricate patterns in smaller datasets[85]. This allowed for a comparative analysis of their predictive performance. Following the ensemble learning methodology outlined in Salehian, et al. [66], an ensemble of DNNs (with 20 models per ensemble in this study) was trained in different random seeds, with the final output for porosity or tensile strength being the average of the individual models' outputs. This ensemble modelling strategy enhances the robustness of the DNN by mitigating the effects of random initialisation on training performance and enabling the estimation of the standard deviation for future predictions – an important measure for assessing the model's prediction quality for new data points. All DNNs were trained using the same dataset and model architecture, which consisted of two hidden layers with 128 units, each followed by a ReLU activation function.

Given the DNN's superior performance over Random Forest (RF) and Support Vector Regression (SVR), the DNNs were retrained incorporating all parameters listed in Table 3, including the crystallographic and particle informatics descriptors [86,87], to examine the impact of considering API crystal structure on prediction accuracy (*Version2*). The comparative analysis of prediction performance between the initial version (*Version1* in Figure S1 in Supporting Information) and expanded input features (*Version2* in Figure 3) demonstrates an improvement in the accuracy of the DNNs following the inclusion of CSD particle data in the input parameters. This enhanced version of the models has been adopted as the final version and will be utilised in the formulation optimisation framework.

Table 3: Input features for the process models.

| ID | Parameter | Size | Descriptor | Source |
|---|---|---|---|---|
| 1 | Mixture true density | 1 | Blend property | Mixture model |
| 2 | Mixture bulk density | 1 | Blend property | Mixture model |
| 3 | PCs of particle size distribution | 3 | Blend property | Mixture model |
| 4 | PCs of aspect ratio distribution | 3 | Blend property | Mixture model |
| 5 | Tapped density | 1 | Blend property | Mixture model |
| 6 | Flowability (FFC) | 1 | Blend property | Mixture model |
| 7 | Main compression pressure | 1 | Process condition | Process settings |
| 8 | API concentration (drug loading) | 1 | Formulation | Formulation |
| 9 | Informatics descriptors* | 9 | Calculated particle properties | CSD Python API [87] |

*The list of informatics descriptors are provided in Table S4 in supporting information.

### 7.2.2 In-silico Optimisation

The in-silico formulation and process utilising the hybrid system of models is defined as:

$$\underset{\substack{x \in \mathbb{R}^{N_x} \\ m \in \mathbb{R}^{N_m}}}{J}(x, m) = -FFC$$

Subject to:

$$\theta_{\hat{\sigma}} - [E(\hat{\sigma}) - \alpha \times \delta_{\hat{\sigma}}] < 0$$

$$\theta_{\hat{\epsilon}} - [E(\hat{\epsilon}) - \beta \times \delta_{\hat{\epsilon}}] < 0$$

Eq. 1

where $x$ is the $N_x$ dimensional vector of decision variables; $m$ is the $N_x$ dimensional state vector of raw component properties (e.g. particle size and aspect ratio distribution, true density, bulk density); $E(\hat{\epsilon})$ and $E(\hat{\sigma})$ are the expected mean value of the predicted porosity and tensile strength, respectively; $\theta$ is the user-defined threshold for processability conditions ($\theta_{\hat{\sigma}} = 2$ MPa and $\theta_{\hat{\epsilon}} = 0.15$ in this study), $\delta_{\hat{\epsilon}}$ and $\delta_{\hat{\sigma}}$ are the standard deviation of the predicted porosity and tensile strength, respectively; $\alpha$ and $\beta$ are user-defined constants (both are set to 0.2 in this study) to define the allowable level of risk in the robust optimisation process. Notably, higher values of these risk factors result in a more conservative optimisation, thereby reducing the likelihood that the formulation will fail validation.

Non-dominated Sorting Genetic Algorithm II (NSGA-II)[88] was used as the optimisation algorithm due to its proven capability in global search and independence from calculating the gradient . The population size and the number of iterations were set to 30 and 50, respectively.

### 7.3 Tableting DataFactory

#### 7.3.1 Hardware

The Tableting DataFactory setup (Figure 5) was built on an M6 tapped table spanning 200 x 200 cm². R1 and R2 has a reach of 850 mm and 820 mm and can carry up to 5 kg and 14 kg of load, respectively. The FlexPTS (DEC Group, Switzerland) technology is used to dose the pre-mixed powder blend. The quantity of the powder discharged from the dosing unit is volumetric based where the volume can be adjusted by setting the height of a piston altering the powder chamber height. The dosing unit collects the powder in the chamber using a vacuum, the powder is discharged using compressed air. The controller of the dosing unit has the capability to change the duration of the vacuum pump and the pressure of the compressed air to dispense powders with different physical properties, e.g. to consider variations in density, and particle size/shape.

The R1 gripper releases the TU on the weighing balance (Cole-Parmer PA-224I, United States) that is placed under the dosing unit. The tube that carries the powder in TU has the capacity of 3000 mm³ (Figure S3 in Supporting Information) A sliding gate operated through a linear solenoid is used to hold and then release the powder dosed from the dosing unit. To perform NIR measurements for blend homogeneity assessment, the gate of the TU incorporates a 1-mm thick sapphire glass window with 10 mm in diameter. The round tube that holds the powder has a diameter of 8 mm. As this small tube can pose a challenge when discharging adhesive and cohesive materials, the inside of the tube was coated with PTFE to create a non-stick surface and a motorised vibrator, which is activated upon the opening of the gate. Two customised 3D printed fingers are mounted on the robotic gripper that grasp the TU from the back to transport the powder to the different stations (Figure S4 in Supporting Information). As the mass of the powder discharged into the TU is weighed prior to tablet manufacturing, the electronic devices in the TU need to be electrically connected with fingers through metallic touchpoints to allow the TU to remain connection-free and standalone when it is placed on the weighing balance. Through this connection, the electric solenoid and the vibrator are operated by an external and customised electronic control unit. The control unit for the TU receives the control commands from the SCU via serial communication and operates the solenoid and vibrator based on an external power supply. The SCU also acquires the initial weight of the TU before getting the dose and subtracts this weight from the final weight to determine the true value of powder obtained in that iteration.

A NIR spectrometer (Micro NIR PAT-W, VIAVI, United States) is incorporated in the workflow to assess blend homogeneity. NIR is a widely adopted technique in the pharmaceutical sector for swift, non-invasive, and non-destructive analysis, without the need for sample preparation [89]. As changing ambient conditions may influence the NIR measurement, new dark and the reference scans need to be acquired over time. Therefore, the 99% reflectance disc is attached to R1 to take the reference scan at the beginning of each iteration.

Tablets are produced using a tablet press (STYL'One Nano, MEDELPHARM, France). The door of the tablet press was replaced by a laser curtain to provide the robotic arm R1 easy access to the compaction die for powder discharge. Control system of the tablet press waits until the laser curtain is interrupted by R1 to fill the die and move away before initiating the powder compaction process. The tablet chute is kept within the boundaries of the laser curtain and linked with the tablet tester through a side wall to avoid any process interruption.

A fully automated tester (AT50, SOTAX, Switzerland) is used to measure breaking force, weight, diameter, and thickness of the tablets. The testing process starts as a tablet enters the feeder of the tester, then moving through various stations to assess its properties. A bespoke tablet separator (TS) is used to differentiate between damaged and undamaged tablets. The flow of tablets is controlled by adjusting the position of a motor-driven barrier (Figure S5 in Supporting Information). To ensure proper alignment of the undamaged tablets, a linear solenoid pushes them onto their flat surface. Bespoke robotic fingers (Figure S5 in Supporting Information) are mounted on the R2's robotic gripper to accurately grip the tablet for transport. The CU is attached to a high-power vacuum cleaner that is operated by an external control unit. Operation of CU is signalled by the SCU, and it only runs when the TU requires cleaning.

The SCU can control and monitor all instruments remotely (Section 3.6 in Supporting Information). Each instrument is assigned a dedicated computer for two main reasons: 1) to provide local access for users who may need to operate the instrument separately for other tasks, and 2) to standardise diverse communication protocols into a unified protocol for seamless communication with the SCU (as illustrated in Figure S8 in Supporting Information).

### 7.3.2 Real-time Process Optimisers

For the PIBO, the compressibility model developed by Kawakita [90,91] and the compactability model developed by Ryshkewitch-Duckworth [92,93] were used as the physics-based models due to their proven capability in capturing the compression profile [27]:

$$\varepsilon(P) = \frac{\varepsilon_0}{1 + (\frac{V_\infty}{V_0})bP} = \frac{\varepsilon_0}{1 + BP} \qquad \text{Eq. 2}$$

$$\sigma(\varepsilon) = \hat{T}e^{-k_b\varepsilon} \qquad \text{Eq. 3}$$

where $\varepsilon$ is the porosity of the tablet, $\sigma$ is the tensile strength of the tablet, and $P$ is the main compression pressure. For the Kawakita model, $\varepsilon_0$ is the initial powder bed porosity, $V_\infty$ is the net volume of powder, $V_0$ is the initial apparent volume of powder, and $b$ is a tuning parameter which is hypothesised to reflect the resistant and cohesive forces of the particles [94]. Following [27], the ratio of volumes $\frac{V_\infty}{V_0}$ and the constant $b$ were grouped into the single tuning parameter $B$ to simplify the fitting process. For Ryshkewitch-Duckworth model, $\hat{T}$ is the strength at zero porosity and $k_b$ represents the material's bonding capacity. The acquisition function in the classic (black-box) Bayesian optimisation is constrained such that:

$$Err(f_K(\underline{X}), \mathcal{D}_n) \leq Err(f_K(\underline{X}), \mathcal{D}_{n-1}) \qquad \text{Eq. 4}$$

$$Err(f_{R-D}(\underline{X}), \mathcal{D}_n) \leq Err(f_{R-D}(\underline{X}), \mathcal{D}_{n-1}) \qquad \text{Eq. 5}$$

where $f_K$ and $f_{R-D}$ are the Kawakita and Ryshkewitch-Duckworth models, respectively; $\mathcal{D}_n$ refers to the dataset of collected experimental observations at iteration $n$; $Err(f(\underline{X}), \mathcal{D}_n)$ denotes the error (root mean squared error in this study) between the physics-based model and the collected data at iteration $n$. These constraints mean that while minimising the acquisition function from black-box BO, the optimisation trajectory is confined to reduce the error between the collected data and physics-based models. The optimisation is terminated if the change in the tuning parameters of physics-based models within the last two iteration falls below a user-defined minimum threshold (in this study, set to be less than 20% change in the tuning parameter as compared to the previous iteration). These criteria were established to ensure the tuning parameters of compressibility and compactability profiles are calculated with sufficient accuracy, enabling reliable prediction of the profile across a continuous range of input parameters.

For the MOBO , a Design of Experiments (DoE) using the Latin Hypercube Sampling (LHS) method was initially employed to generate a diverse set of experimental conditions to pre-train the process models [95]. LHS is used due to its space-filling features to generate training datasets that are evenly distributed over the design space to ensure good coverage. This was followed by a classic black-box optimisation approach to iteratively refine the decision parameters space. Three independent GP models were trained during MOBO to individually predict elastic recovery, porosity, and tensile strength based on the input parameters. The target was set to minimise the elastic recovery while meeting the porosity ($\varepsilon \geq 0.15$) and tensile strength ($\sigma \geq 2$ MPa) constraints. The LHS was run for 15 experiments followed by 25 iterations of Bayesian optimisation. The optimisation process was terminated after the predetermined number of experiments.

### 7.4 Extended Reality

An immersive digital twin of the Tableting DataFactory was implemented using a real-time development platform (Unity) targeting Microsoft's HoloLens 2 mixed reality devices. 3D models of the Tableting DataFactory equipment were constructed with Autodesk's 3D Studio Max using 3D CAD files, brochures, photos and measurements taken on site. The 3D lab models were then imported into the Unity environment for uploading to the HoloLens devices. To gather the data from the Tableting DataFactory, a REST API (allowing systems to exchange data over the internet) was implemented to take advantage of the integration capabilities of LabVIEW. The HoloLens can access the experimental data in real-time as long as it is connected online. This enables the user to view real-time experimental data using either the AR digital twin at the lab (data overlaid onto real lab equipment), or the MR digital twin (laboratory hologram) while working remotely.

The AR version of the digital twin aligns all the overlaid data onto the associated equipment within the Tableting DataFactory. This is achieved using a single QR code that the HoloLens detects and uses as a reference to calibrate the entire scene.

The MR version of the digital twin can be used in any location. It uses surface (floor) detection enabling the user to place a scalable Tableting DataFactory hologram into any suitable location.

The digital twin enables users to receive real-time updates to the lab's holographic representation and data overlays, even when they are located remotely from the physical lab.

**Authorship Contribution Statement**

**Mohammad Salehian and Faisal Abbas** (equally contributed)**:** Conceptualisation, Data curation, Formal analysis, Investigation, Methodology, Software, Validation, Visualisation, Writing – original draft. **Peter Hou:** Conceptualisation, Investigation, Methodology. **Jonathan Moores:** Data curation, Methodology, Writing – review & editing. **Jonathan Goldie:** Data curation, Methodology. **Alexandros Tsioutsios:** Data curation, Writing – review & editing. **Victor Portela:** Software, Visualisation, Data curation, Writing – review & editing. **Quentin Boulay:** Resources, Writing – review & editing. **Roland Thiolliere:** Resources, Writing – review & editing. **Ashley Stark:** Resources, Writing – review & editing. **Jean-Jacques Schwartz:** Resources, Writing – review & editing. **Jerome Guerin:** Resources, Writing – review & editing. **Andrew G. P. Maloney:** Data curation, Resources, Writing – review & editing. **Alexandru A. Moldovan:** Data curation, Resources, Writing – review & editing. **Gavin K. Reynolds:** Resources, Writing – review & editing. **Jérôme Mantanus:** Resources, Writing – review & editing. **Catriona Clark:** Resources, Writing – review & editing. **Paul Chapman:** Resources, Supervision, Writing – review & editing. **Alastair Florence:** Funding acquisition, Supervision, Writing – review & editing. **Daniel Markl:** Conceptualisation, Funding acquisition, Methodology, Project administration, Supervision, Writing – review & editing.

**Acknowledgements**

The authors would like to thank the Digital Medicines Manufacturing (DM2) Research Centre Co-funded by the Made Smarter Innovation challenge at UK Research and Innovation (Grant Ref: EP/V062077/1) for funding this work. The authors also thank the EPSRC ARTICULAR project (Grant ref: EP/R032858/1) and EPSRC Future Continuous Manufacturing and Advanced Crystallisation Research Hub (Grant Ref: EP/P006965/1) for the data generated and exploited in this work This work was also supported by Research England and the Scottish Funding Council under the UK Research Partnership Investment Fund Net Zero Medicines Manufacturing Research Pilot.

# Supporting Information for "Accelerated Medicines Development using a Digital Formulator and a Self-Driving Tableting DataFactory"

Faisal Abbas[1,^] & Mohammad Salehian[1,^], Peter Hou[1], Jonathan Moores[1], Jonathan Goldie[1], Alexandros Tsioutsios[1], Victor Portela[2], Quentin Boulay[3], Roland Thiolliere[3], Ashley Stark[4], Jean-Jacques Schwartz[4], Jerome Guerin[4], Andrew G. P. Maloney[5], Alexandru A. Moldovan[5], Gavin K. Reynolds[6], Jérôme Mantanus[7], Catriona Clark[1], Paul Chapman[2], Alastair Florence[1], Daniel Markl[1*]

[^] Authors contributed equally.
[*] Corresponding author, Email: daniel.markl@strath.ac.uk

[1] Centre for Continuous Manufacturing and Advanced Crystallisation (CMAC), Strathclyde Institute of Pharmacy and Biomedical Science (SIPBS), University of Strathclyde, Glasgow, G1 1RD, UK
[2] Glasgow School of Art, Glasgow, G3 6RQ, UK
[3] Medelpharm, ZAC des Malettes, 615 Rue du Chat Botté, 01700 Beynost, France
[4] DEC Group, Chemin du Dévent 3, 1024 Ecublens, Switzerland
[5] The Cambridge Crystallographic Data Centre, 12 Union Road, Cambridge, CB2 1EZ, UK
[6] Sustainable Innovation & Transformational Excellence (xSITE), Pharmaceutical Technology & Development, Operations, AstraZeneca UK Limited, Macclesfield, SK10 2NA, UK
[7] UCB S.A, 60 Allée de la Recherche, 1070 Brussels, Belgium

# 1 Materials and Formulations

Four distinct quinary blends (B1 to B4), exhibiting varying concentrations of paracetamol at 16%, 18%, 20% and 22%, and 20% w/w, respectively, were formulated as presented in Table S1. All the blends were prepared using a laboratory powder blender (PharmaTech Multiblend MB015). The blending process was performed at a blender speed of 20 rpm and an agitator speed of 200 rpm over a duration of 20 minutes to ensure homogeneity. In addition to the formulation process, all blends were lubricated with 1% w/w of magnesium stearate, a procedure implemented to improve the flow properties of the blends. This lubrication step was executed by allowing the blend to mix for an additional 5 minutes, thus ensuring a uniform distribution of the lubricant without overlubricating the mixture. In our study, we also prepared five quinary blends (B5 to B9) using different APIs, as illustrated in Table S2, each planned to include a fixed concentration of the API (20% w/w), croscarmellose sodium (3.5% w/w) and magnesium stearate (1%). These blends were designed to have varied filler ratios to broaden the knowledge space. The filler combinations utilised included different grades of microcrystalline cellulose and lactose. The combination and concentrations of fillers were optimized using models. Consistency was maintained in the formulation process across the blends, employing the same laboratory powder blender (PharmaTech Multiblend MB015) with fixed settings of a 20-rpm blender speed and a 200-rpm agitator speed, operating over a duration of 20 minutes. A detailed overview of the precise formulations and the corresponding concentrations (% w/w) of each filler in each blend prepared are illustrated in Table S1.

Table S1: Characteristics of excipients and active pharmaceutical ingredient used in the tablet formulations. Information includes the material grade, and supplier, and physical properties like Bulk, Tapped, and True Density (measured in mg/mL), the Flow Function Coefficient (FFC), and volume-based particle size percentiles (d10, d50, d90, in μm).

| Material (ID) | Grade | Supplier | Bulk, Tapped, True Density (mg/mL) | FFC @ KPa | d10, d50, d90 VB (μm) |
|---|---|---|---|---|---|
| Paracetamol (SP) | Standard 6375 | Mallinckrodt | 0.34, 0.55, 1.18 | 1.04 @ 0.30 | 24.47, 88.18, 205.00 |
| Aspirin (AS) | Powder | Molekula | 0.77, 0.86, 1.394 | 9.29 @0.81 | 68.76, 148.2, 322.7 |
| Dexamethasone (DM) | Powder | Molekula | 0.322, 0.411, 1.385 | 1.34 @0.80 | 9.73, 25.18, 176.6 |
| Griseofulvin (GR) | Powder | Molekula | 0.36, 0.52, 1.58 | 2.00 @0.29 | 14.47, 29.67, 99.01 |
| Indomethacin (IM) | Powder | Molekula | 0.36, 0.52, 1.58 | 2.98 @0.29 | 45, 93.02, 165.2 |
| Metformin Hydrochloride (MH) | Powder | Molekula | 0.67, 0.764, 1.347 | 3.91 @0.80 | 22.23, 192, 352.5 |
| Microcrystalline Cellulose (MCC1) | Avicel®PH-101 | DuPont Nutrition | 0.33, 0.45, 1.56 | - | 44.51, 86.33, 153.20 |
| Microcrystalline Cellulose (MCC2) | Avicel®PH-102 | DuPont Nutrition | 0.34, 0.44, 1.56 | 3.91 @0.79 | 27.5, 53.48, 104 |
| Microcrystalline Cellulose (MCC3) | Microcel 302 | Roquette | 0.45, 0.51, 1.53 | 4.44 @0.80 | 25.46, 62.36, 152 |
| Lactose (LAC1) | Fastflo®316 | Foremost Farms USA | 0.63, 0.74, 1.54 | 5.55 @1.06 | 55.10, 104.70, 188.40 |
| Lactose (LAC2) | Granulac 200M | Meggle pharm | 0.52, 0.76, 1.55 | - | 22.42, 59.67, 107.30 |
| Manitol (MAN) | Pearlitol 200 SD | Roquette | 0.54, 0.59, 1.48 | 12.07 @1.06 | 68.78, 115.10, 191.10 |
| Dibasic Calcium Phosphate (DCPA) | Anhydrous Emcompress | JRS Pharma | 0.71, 0.87, 2.98 | 8.27 @0.80 | 77.81, 203.1, 281.8 |
| Croscarmellose Sodium (CCS) | AcDiSol | FMC International | 0.54, 0.74, 1.60 | 3.6 @0.80 | 30.39, 51.58, 82.14 |
| **Magnesium Stearate (MgSt)** | Hyqual 5712 | Mallinckrodt | 0.28, 0.35, 1.30 | 1.44 @0.30 | 7.02, 16.12, 41.01 |

Table S2: Composition of the tablet blends used for benchmarking and validation of Tableting DataFactory. The data in the table represents the percentage weight by weight (w/w) of each ingredient in the formulation. The blend ID describes the unique identifier for each blend, followed by the concentration of disintegrant, filler 1, filler 2, and lubricant in the blend, respectively

| Blend ID | API (% w/w) | Disintegrant (% w/w) | Filler 1 (% w/w) | Filler 2 (% w/w) | Lubricant (% w/w) |
|---|---|---|---|---|---|
| **B1** | SP (16%) | CCS (3.5%) | LAC1 (30.4%) | MCC2 (49.1%) | MgSt (1%) |
| **B2** | SP (18%) | CCS (3.5%) | LAC1 (23%) | MCC2 (54.5%) | MgSt (1%) |
| **B3** | SP (20%) | CCS (3.5%) | LAC1 (2.7%) | MCC2 (72.8%) | MgSt (1%) |
| **B4** | SP (22%) | CCS (3.5%) | - | MCC2 (73.5%) | MgSt (1%) |
| **B5** | AS (20%) | CCS (3.5%) | LAC1 (27.7%) | MCC2 (47.8%) | MgSt (1%) |
| **B6** | DM (20%) | CCS (3.5%) | MCC1 (57.3%) | MCC2 (18.2%) | MgSt (1%) |
| **B7** | GR (20%) | CCS (3.5%) | LAC1 (41.3%) | MCC3 (34.2%) | MgSt (1%) |
| **B8** | IM (20%) | CCS (3.5%) | LAC1 (57.3%) | MCC2 (18.2%) | MgSt (1%) |
| **B9** | MH (20%) | CCS (3.5%) | LAC1 (30.5%) | MCC2 (45%) | MgSt (1%) |
| **B10** | SP (1%) | - | LAC1 (98%) | - | MgSt (1%) |
| **B11** | SP (5%) | - | LAC1 (94%) | - | MgSt (1%) |
| **B12** | SP (10%) | - | LAC1 (89%) | - | MgSt (1%) |
| **B13** | SP (15%) | - | LAC1 (84%) | - | MgSt (1%) |
| **B14** | SP (20%) | - | LAC1 (79%) | - | MgSt (1%) |
| **B15** | - | CCS (5%) | MAN (30%) | DCPA (64%) | MgSt (1%) |
| **B16** | - | CCS (5%) | MAN (30%) | LAC1 (64%) | MgSt (1%) |
| **B17** | - | CCS (5%) | MAN (64%) | LAC1 (30%) | MgSt (1%) |
| **B18** | - | CCS (5%) | MCC (30%) | DCPA (64%) | MgSt (1%) |
| **B19** | - | CCS (5%) | MCC (64%) | DCPA (30%) | MgSt (1%) |

## 2 Digital Formulator
### 2.1 Hybrid System of models

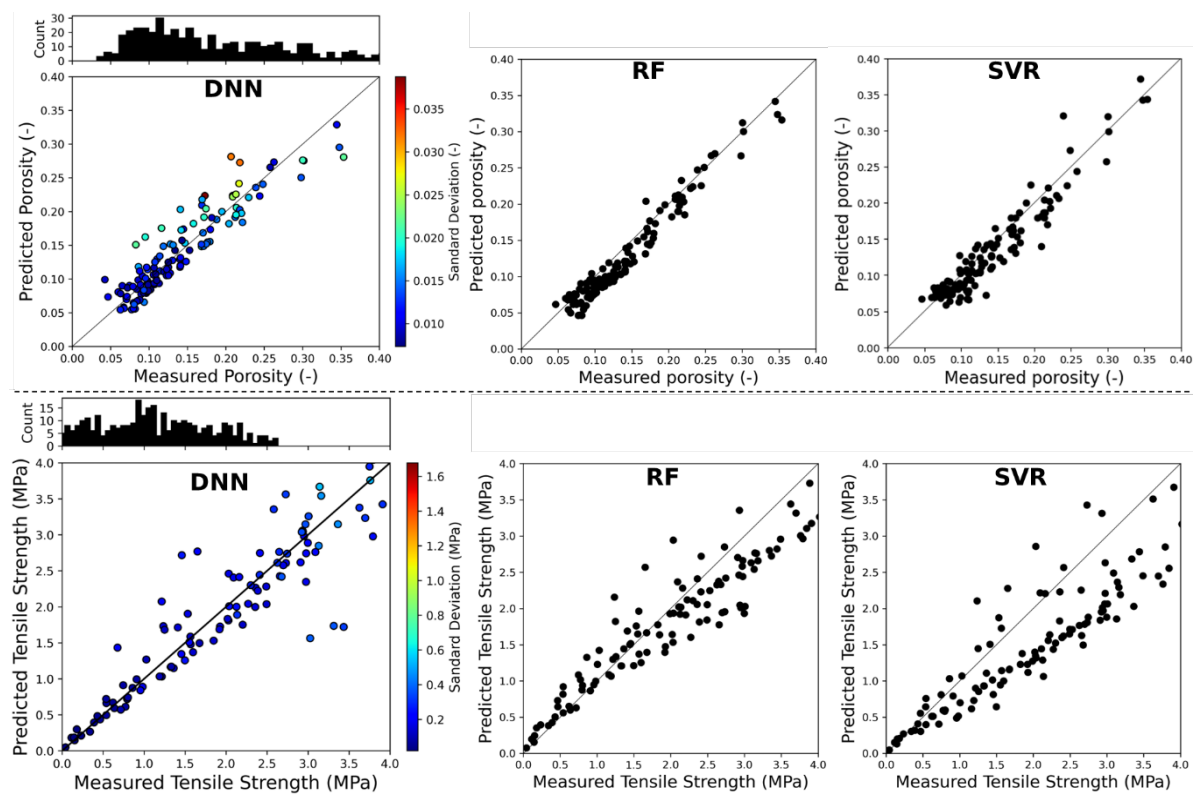

Figure S1: Prediction performance of (top) porosity and (bottom) tensile strength models using DNN, RF, and SVM. Particle informatics descriptors are excluded from input parameters. The histograms on top of DNN show the frequency distribution of training data. The colour bars show the estimated standard deviation of predicted test data.

Table S3: Summary of tablet data

|  | Drug loading (-) | | Porosity (-) | | Tensile Strength (MPa) | | Main Compression Pressure (MPa) | |
| --- | --- | --- | --- | --- | --- | --- | --- | --- |
| API | Min. | Max. | Min. | Max. | Min. | Max. | Min. | Max. |
| **Placebo** | 0 | 0 | 0.032 | 0.386 | 0.031 | 13.916 | 5.533 | 565.004 |
| **SP** | 0.010 | 0.315 | 0.043 | 0.404 | 0.041 | 8.177 | 12.276 | 489.029 |
| **GR** | 0.300 | 0.300 | 0.095 | 0.173 | 3.026 | 3.435 | 157.190 | 157.190 |
| **BZ** | 0.350 | 0.350 | 0.063 | 0.811 | 0.874 | 4.519 | 54.696 | 398.001 |
| **LOV** | 1.000 | 1.000 | 0.064 | 0.432 | 0.066 | 2.294 | 12.051 | 326.327 |
| **IBU** | 0.010 | 0.529 | 0.028 | 0.363 | 0.043 | 3.093 | 11.647 | 369.397 |
| **MF** | 0.050 | 0.464 | 0.063 | 0.375 | 0.029 | 4.481 | 13.801 | 367.573 |

Table S4: List of crystal structure and Particle Informatics descriptors used for data-driven prediction of porosity and tensile strength.

| API | Packing Coefficient | Average Hydrogen Bond Donor Density | Average Hydrogen Bond Acceptor Density | Average Rugosity | Average Surface Charge | Short-to-Medium Axis Length Ratio | Medium-to-Long Axis Length Ratio | Hydrogen Bond Dimensionality |
|---|---|---|---|---|---|---|---|---|
| SP | 0.727 | 0.046 | 0.043 | 1.826 | -0.196 | 0.868 | 0.835 | 2 |
| GR | 0.703 | 0 | 0.069 | 1.659 | -0.356 | 1 | 0.554 | -1 |
| IBU | 0.676 | 0.009 | 0.018 | 1.768 | -0.212 | 0.424 | 0.864 | 0 |
| IM | 0.703 | 0.012 | 0.055 | 1.643 | -0.479 | 0.697 | 0.425 | 0 |
| BZ | 0.713 | 0.026 | 0.053 | 1.265 | -0.302 | 0.478 | 0.927 | 0 |
| LOV | 0.702 | 0.005 | 0.033 | 1.547 | -0.058 | 0.864 | 0.467 | 1 |
| MF | 0.691 | 0.040 | 0.061 | 1.729 | -0.372 | 0.454 | 0.838 | 0 |

Table S5: Validation accuracy ($R^2$ and RMSE) of process models with and without including CSD parameters.

| | Without CSD parameter | | | | With CSD parameters | | | |
|---|---|---|---|---|---|---|---|---|
| | Porosity | | Tensile Strength | | Porosity | | Tensile Strength | |
| | R2 | RMSE | R2 | RMSE | R2 | RMSE | R2 | RMSE |
| **DNN** | 0.90 | 0.024 | 0.86 | 0.28 | 0.95 | 0.019 | 0.90 | 0.25 |
| **RF** | 0.89 | 0.031 | 0.72 | 0.40 | - | - | - | - |
| **SVM** | 0.83 | 0.048 | 0.67 | 0.44 | - | - | - | - |

## 2.2 In-silico Optimisation

Table S6: List of decision parameters and their values/ranges in the formulation optimisation cases.

| Decision parameter | Value/Range |
| --- | --- |
| API (-) | SP, AS, DM, GR, IM, MH |
| API mass fraction (-) | 0.16, 0.18, 0.2, 0.22 |
| Excipient 1 (-) | Choice from MCC (3 grades), LAC (2 grades), MAN (1 grade) |
| Excipient 1 mass fraction (-) | [0 – 1] |
| Excipient 2 (-) | Choice from MCC (3 grades), LAC (2 grades), MAN (1 grade) |
| Excipient 2 mass fraction (-) | [0 – 1] |
| Lubricant (-) | MgSt |
| Lubricant mass fraction (-) | 0.035 |
| Disintegrant (-) | CCS |
| Disintegrant mass fraction (-) | 0.01 |
| Compression Pressure (MPa) | [70 – 450] |

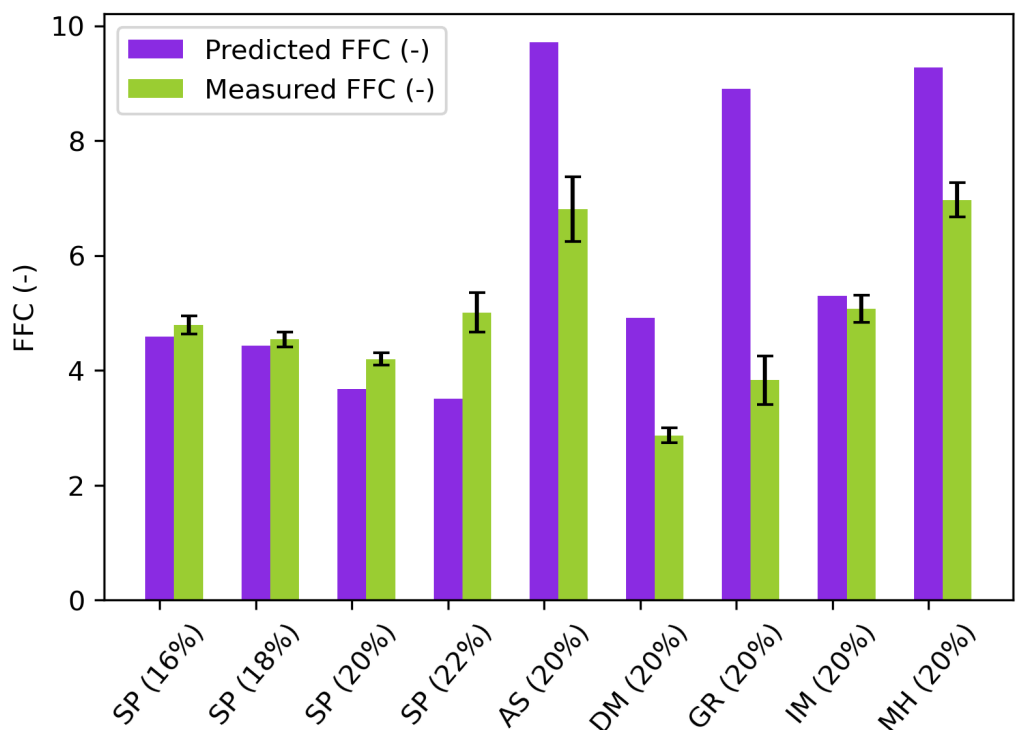

Figure S2: Validation of FFC values against the measured data at consolidation pressure of 1.6 KPa. The error bars show the standard deviation from multiple measurements.

# 3 Tableting DataFactory
## 3.1 List of instruments

Table S7: List of all the instruments used and their digital interfaces

| Process | Instrument | Digital interface |
|---|---|---|
| Powder dosing | DEC Flex PTS | Profinet |
| NIR spectroscopy | VIAVI Micro NIR | OPC UA |
| Powder compaction | MEDELPHARM STYL'One Nano | Websocket |
| Tablet tester | Sotax AT50 | OPC DA |
| Robot 1 | Universal UR5e | TCP/IP |
| Robot 2 | Kuka LBR iiwa 14 | UDP |
| Weighing balance | Cole-parmer | Serial Communication |

## 3.2 Powder transportation unit (TU)

The custom-built transportation unit consists of a tube that contains powder for transport and analysis as shown in Figure S3. The tube's inlet has a width of 20 mm, allowing sufficient space for powder dosing. The outlet, however, is 8 mm wide, which is just below the 9 mm diameter of the tablet press die where the powder will be discharged. Powder flow is regulated by a sliding gate driven by an electronic solenoid, while an electric vibrator ensures smooth powder flow through the tube. A clear sapphire disk, 10 mm in diameter, is placed above the sliding gate to allow for NIR scanning. A robotic arm (R1) performs pick-and-place tasks by holding the transportation unit from the back and supplying power to the vibrator and solenoid.

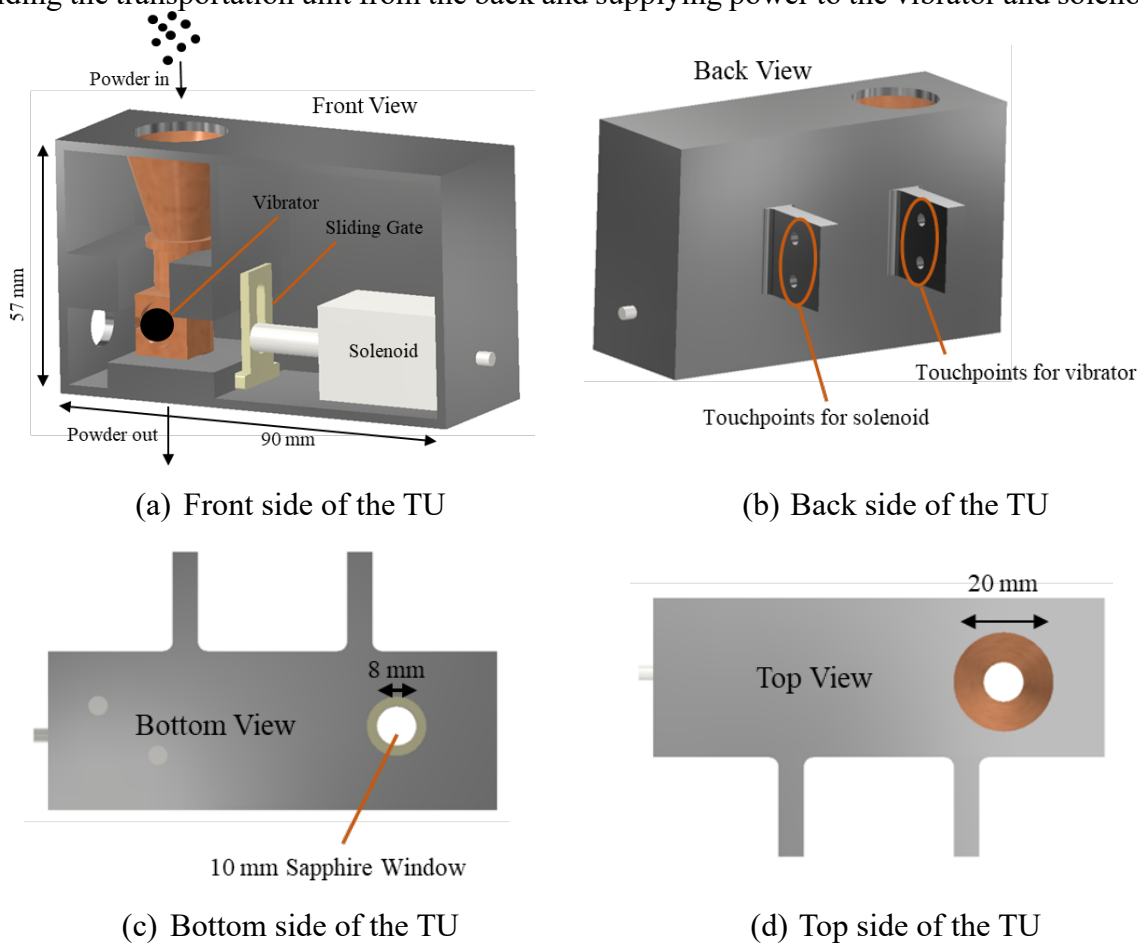

(a) Front side of the TU  (b) Back side of the TU

(c) Bottom side of the TU  (d) Top side of the TU

Figure S3: Design of transportation unit (TU)

## 3.3 Robotic Fingers

Figure S4 shows the design of customized 3D-printed fingers mounted on R1, which enable both the picking and placing of the TU as well as the provision of electrical energy through touchpoints. Figure S5 illustrates the design of customized fingers for R2, which are specifically made for tablet transportation. The fingers are engineered to securely pick up tablets from both horizontal and vertical orientations, ensuring that the tablets remain securely held without slipping.

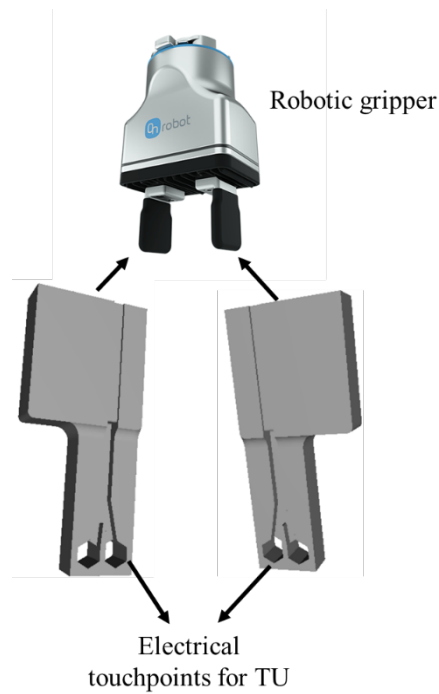

Figure S4: Fingers for the gripper of R1 to hold TU and to provide electrical power

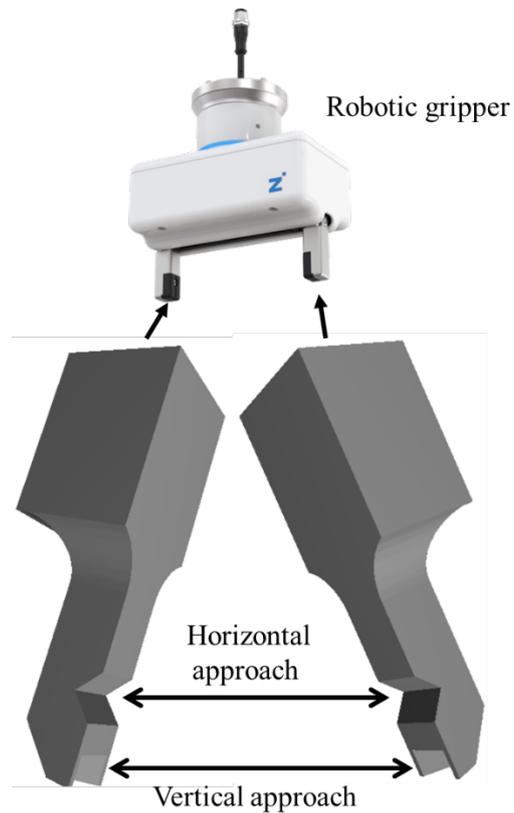

Figure S5: Fingers for the gripper of R2 to pick and place tablets

### 3.4 Automated Tablet Separator

The Automated Tablet Separator (ATS), illustrated in Figure S6, is designed to separate damaged tablets from undamaged ones. The ATS unit directs tablets into two distinct channels, with the flow into each channel regulated by a barrier driven by a servo motor. When an undamaged tablet arrives in the designated channel, a linear solenoid pushes it forward, making it more accessible for a robotic gripper to pick up. Position of the barrier to separate the damaged and undamaged tablet is determined based on the method (destructive/non-destructive) that was used to test the tablets.

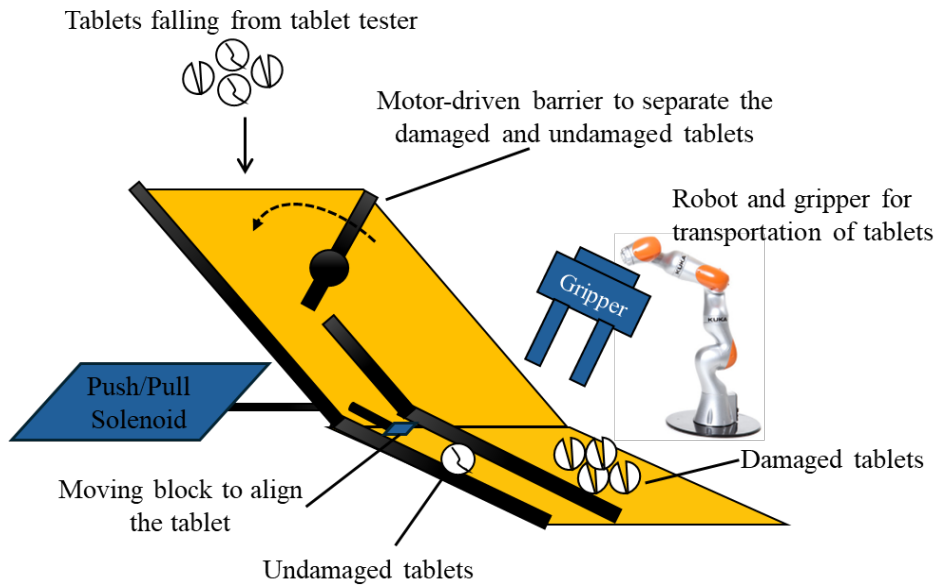

Figure S6: Tablet separator for damaged and undamaged tablets

### 3.5 Cleaning unit for TU

The customized cleaning unit (CU) is designed to clean the TU from front, top and bottom from any powder residue. This CU is attached to a high-powered vacuum cleaner that is operated by an external control unit as shown in **Figure S7**. Operation of this CU can be fully controlled from LabVIEW in real-time.

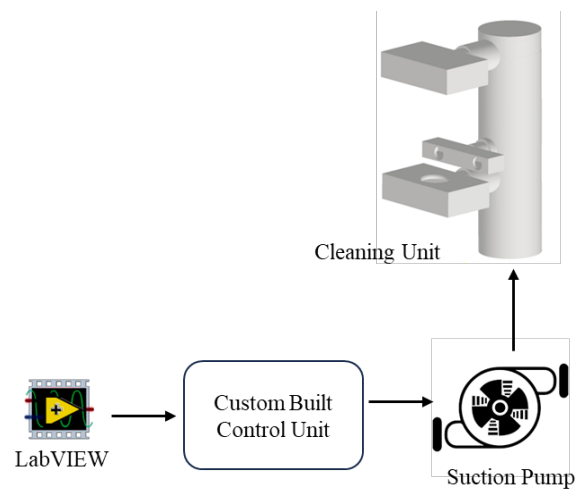

Figure S7: Design of customised CU

### 3.6 Digital Integration of Instruments

Figure S8 shows how all the devices and instruments are digitally integrated from end to end with supervisory control unit. The dosing unit has its own dedicated programmable logic controller (PLC) and does not require additional software for control. It communicates with the supervisory control unit (SCU) via the Profinet protocol, with each PLC tag (parameters/variables) predefined in the SCU for control purposes. The weighing balance is integrated with the SCU through serial communication. The SCU sends a read command to the balance, which then returns the current weight of the dose. The Micro NIR instrument is controlled via proprietary software (VIAVI Micro NIR), which offers remote access through the OPC interface. All configurations and method-related information are set up in this software. Once the method is created, the dark, reference, and sample scans can be managed through the SCU, with spectra acquired via OPC communication protocol. The tablet press is operated using proprietary Alix software, which communicates with a remote computer through a WebSocket interface. The tablet press's communication interface allows for greater flexibility in adjusting parameters remotely without relying on its proprietary software. Similarly, the tablet tester is managed through Q-doc, a proprietary software installed on a device-specific laptop. All necessary parameters for creating destructive and non-destructive profiles are predefined in Q-doc, which can send and receive data remotely through an OPC interface. The R1 and R2 robots are integrated with the SCU via TCP and UDP socket interfaces, respectively. R1's control system uses block-based programming, where each block represents a specific function, while R2 is programmed using Java-based scripting. Both robots move between instruments based on pre-defined waypoints, with their movements triggered by instructions from the SCU. Finally, all the bespoke units, including TU, CU and TS are integrated with SCU through embedded controller and serial communication. The embedded controller receives the specific instruction from SCU and then performs the actions on bespoke units.

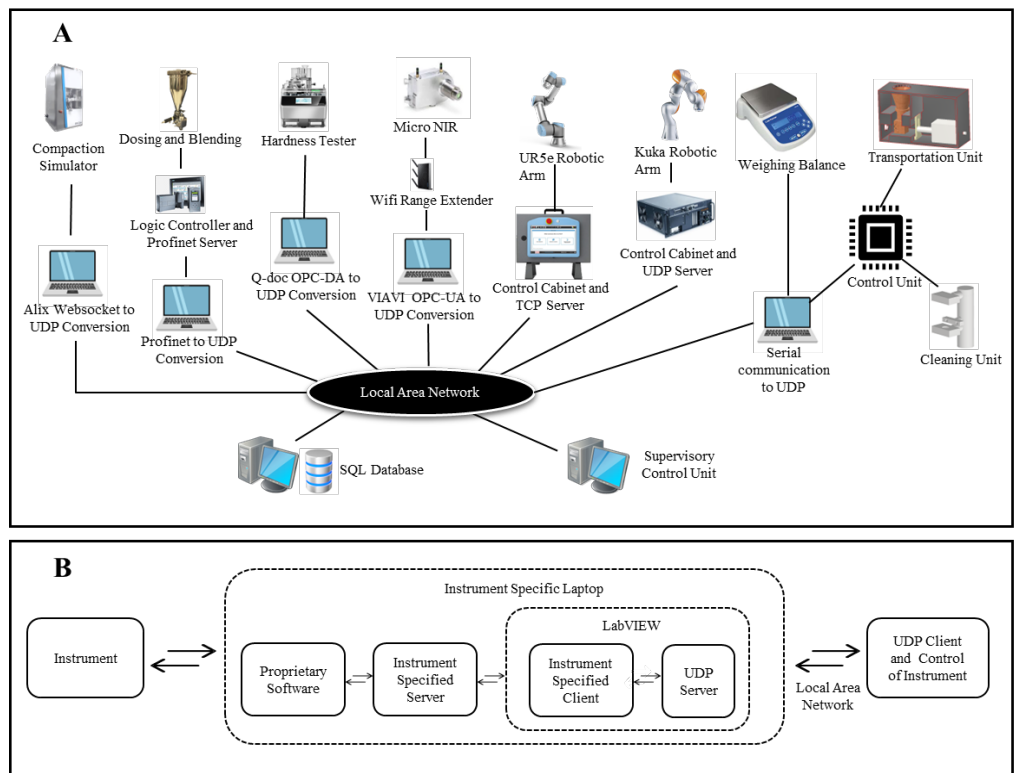

Figure S8: Digital integration of instruments with SCU.

## 3.7 Validation of Tableting DataFactory

Figure S10 (a) reveals that powder loss during transportation, encompassing powder dosing, adherence to the TU tube, and spillage when opening the TU gate, remains consistently below 22 mg for various formulations. Despite the challenges posed by poor flowability in some Paracetamol formulations, TU maintains a consistent powder loss across various formulations. The formulations B6 and B8 exhibit the least powder loss due to their suboptimal flowability, resulting in minimal spillage during powder dosing and release from TU. Figure S10 (b) indicates that there is no significant influence of dose weights on powder loss, with the powder loss consistently staying below 22 mg for all dose weights.

In Figure S11 (a), the repeatability of both powder dose and tablet weight is depicted for different formulations. The standard deviation in the powder obtained from the dosing unit directly corresponds to variations in tablet weight data. As shown in Figure S11 (b), the relative standard deviation in the powder data and tablet weight decreases as the dose weight increases, indicating that the dosing system performs more effectively with higher dose weights.

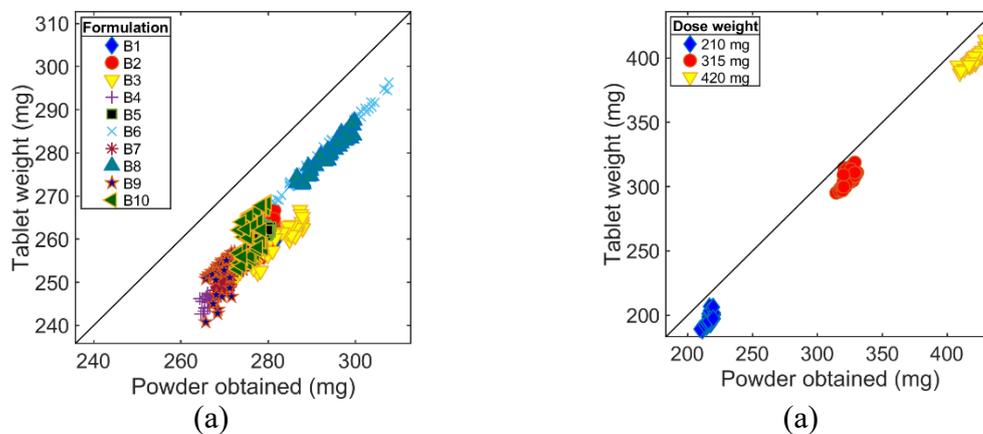

(a)             (a)

Figure S9: Comparison of tablet weight with powder obtained with (a) different formulations (b) different dose weights.

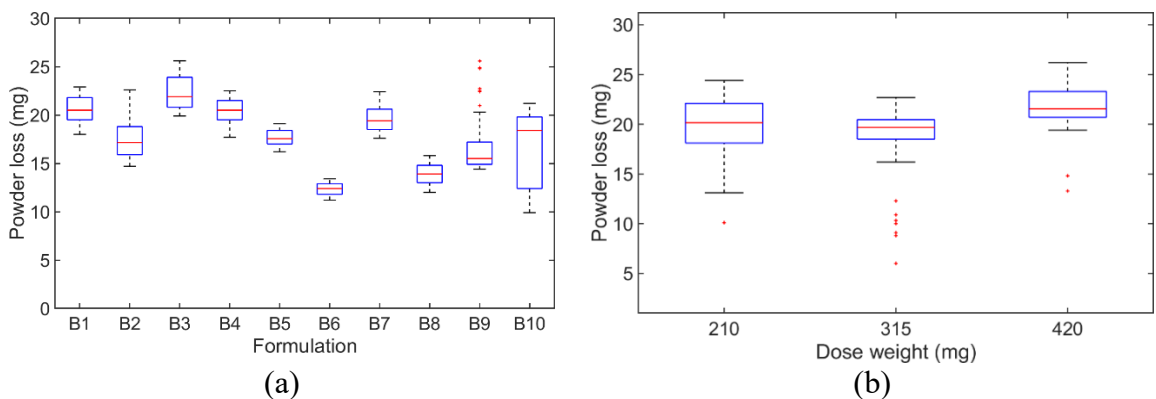

(a)             (b)

Figure S10: Assessment of powder loss in dosing, transportation, and release with (a) different formulations (b) different dose weights

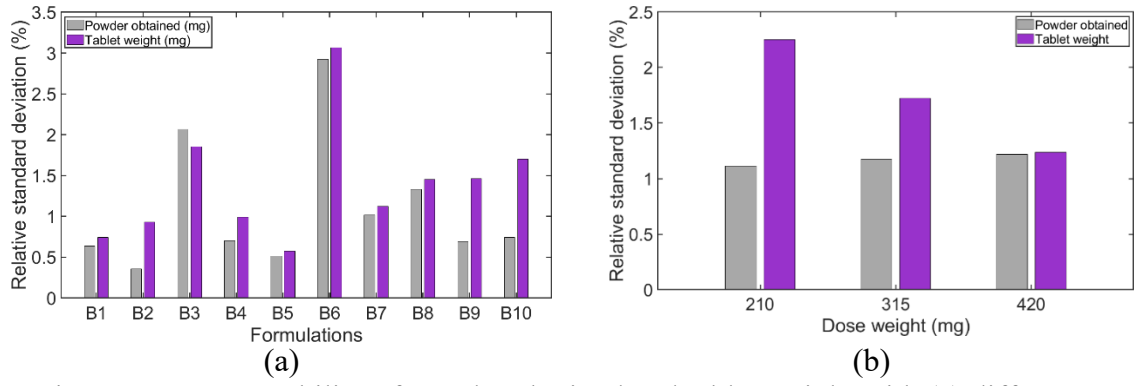

Figure S11: Repeatability of powder obtained and tablet weight with (a) different formulations (b) different dose weights

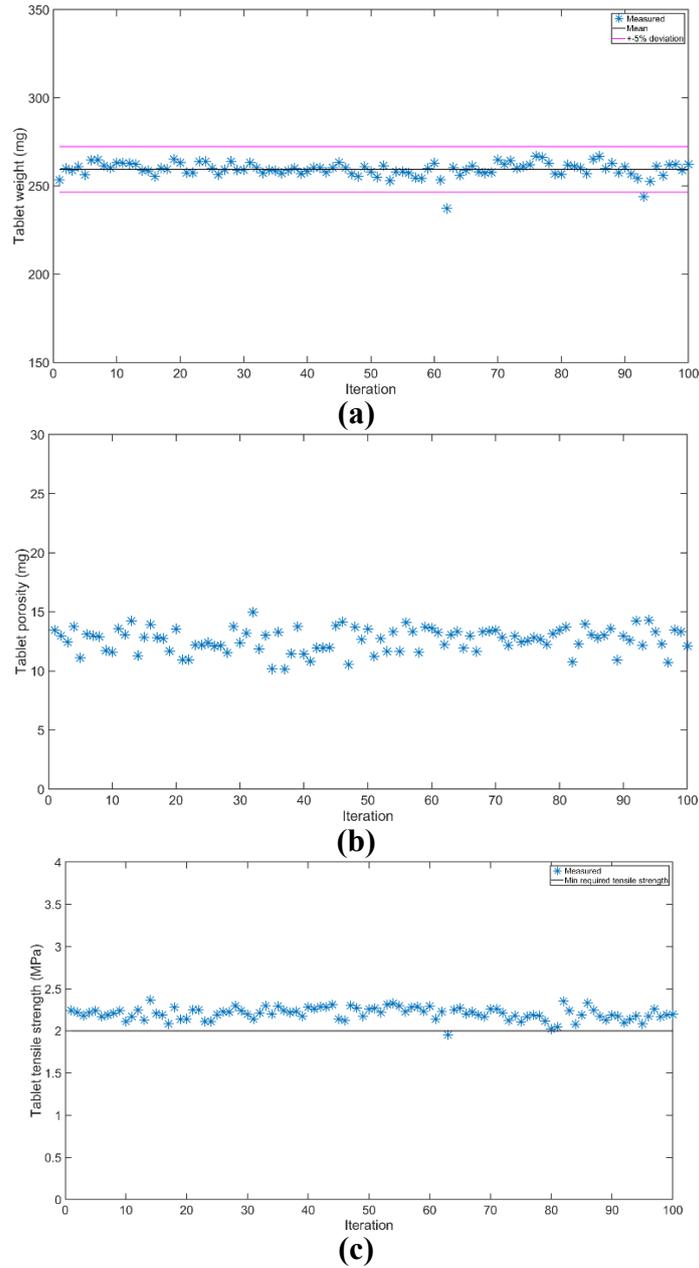

Figure S12: Assessment of variability in tablet a) weight b) porosity c) tensile strength over time.

### 3.7.1 Assessment of Powder Loss on Content Uniformity

Blends were prepared in a Turbula T2GE mixer to achieve formulations containing 5, 6, 7, 7.5, 8, 9, 10, 12.5, 15, 20 and 25 % w/w paracetamol (APAP), alongside 5% croscarmellose sodium (CCS), 1% magnesium stearate (MgSt), and 50:50 mixture of microcrystalline cellulose (MCC) and lactose (LAC) as the remaining excipients making up the %. Pre-lubrication mixing settings were 50 rpm for 10 minutes. Lubrication mixing settings were 50 rpm for 2.5 minutes. Each blend was then subsampled in duplicate for High Performance Liquid Chromatography (HPLC) analysis. The samples were analysed on a Thermo Fisher Vanquish Core system using an Agilent Zorbax Eclipse Plus C18 column (250 mm x 4.6 mm, 5 μm) maintained at 30°C. The mobile phase consisted of 30% acetonitrile and 70% water (v/v) delivered at a flow rate of 1.0 mL/min, and the total run time was set to 10 minutes. Each sample was dissolved, filtered, and diluted (working concentration: 0.15 mg/mL) appropriately. The injection volume set on the instrument method was 2 μL, the autosampler temperature was held at 5°C and the detection wavelength was set to 243 nm.

Following HPLC analysis, five subsamples of each blend were processed through the automated tablet development DataFactory. Each subsample was dosed in the 3D-printed Transportation Unit (TU), analysed by Near InfraRed Spectroscopy (NIRS), and then compacted into a tablet. Each tablet was then subsequently analysed again by NIRS using a similar 3D-printed Unit which enabled accurate tablet placement.

The blend NIR data were pre-processed by wavelength trimming (1100-1450 nm), standard normal variate (SNV) transformation and smoothed via a Savitzky-Golay (SG) second derivative (polynomial order 2, window length 5). A partial least squares model (LV = 2) was produced to correlate the spectra with the reference HPLC measurements. K-fold cross-validation, each fold containing five samples, was applied to evaluate predictive performance, see Figure S13A.

The collected NIR spectra for both blends and tablets were subjected to wavelength trimming (1100-1450 nm), SNV transformation and smoothed via a SG second derivative (polynomial order 2, window length 5). Direct standardisation (DS) was then applied to correct blend spectra into the tablet spectral domain. For each fold in a 5-fold cross-validation set up, a subset of blend-tablet pairs was chosen as the training set, and the remaining pairs formed the test set. A least squares procedure was applied to the training set to solve:

$$\mathbf{X}_{\text{tablet}} = \mathbf{X}_{\text{blend}} \cdot \mathbf{A}$$

where, $\mathbf{X}_{\text{blend}}$ and $\mathbf{X}_{\text{tablet}}$ represent the blend and tablet spectra, respectively, and $\mathbf{A}$ is the transformation matrix that corrects the blend spectra to more closely resemble the tablet domain. The matrix $\mathbf{A}$ computed was then multiplied by the corresponding blend spectra in both training and test sets to produce DS-corrected blend spectra.

Following this, a partial least squares (PLS) regression model (LV =2) was fitted on each training fold using the DS corrected blend spectra and the average HPLC measurements from the original blends. The model was then applied to the DS corrected test spectra to predict content uniformity, yielding predictions which were compared to the HPLC measurements. After collection of all 5-fold predictions, every five subsamples corresponding to the same blend were grouped to measure the mean predicted values and standard deviations (Figure S13B).

The DS corrected outputs ($R^2$ = 0.92, RMSE-CV = 2.01) indicate a robust correlation and low error, comparable to those obtained using the original blend NIR spectra ($R^2$ = 0.97, RMSE-CV = 0.84), as illustrated in Figure S13. Despite unavoidable uncertainties introduced by sample preparation for HPLC, HPLC instrument variability, differences in NIR sampling of powders versus tablets, prediction error from PLS regression, and direct standardization (DS) errors, these findings strongly suggest that any powder losses during processing did not alter the overall composition of the blends. Material was lost in bulk rather than selectively, the ratio of paracetamol to excipients remained stable throughout the process.

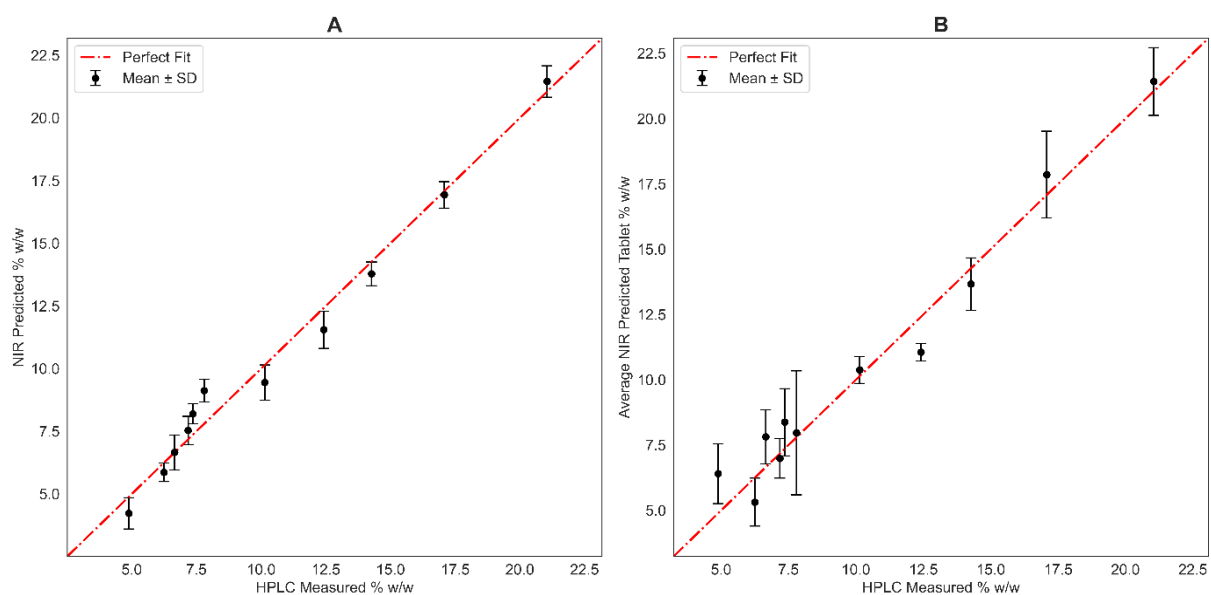

Figure S13: Depicted are the NIR PLS-CV regression results using (A) the original blend spectra and (B) the DS corrected blend spectra targeting the tablet domain. Average NIR predicted vs. HPLC measured paracetamol content (% w/w) for blend samples processed through the automated tablet development DataFactory. Each data point represents the mean of five replicate NIR predictions, with error bars indicating standard deviation. The red dashed line indicates the perfect one to one relationship between NIR predicted and HPLC measured values. Notably, obtaining individual HPLC references for each subsample would further enhance PLS regression accuracy.

### 3.8 Process Analysis with Near-infrared (NIR) Spectroscopy
#### 3.8.1 Background
Near-infrared (NIR) spectroscopy is a rapid, non-destructive analytical technique and widely used for process monitoring and control in pharmaceutical applications. NIR spectra capture key chemical and physical properties of samples, providing valuable information about many critical parameters, such as blend homogeneity. However, raw NIR spectra often suffer from baseline shifts, scattering effects, and noise, necessitating robust pre-processing and dimensionality reduction techniques to ensure reliable analysis.

This study incorporates trimming, standard normal variate (SNV) transformation, and Savitzky-Golay (SG) filtering for pre-processing. Principal component analysis (PCA) is used for dimensionality reduction, while Hotelling's T² analysis is employed for outlier detection and quality assessment of the spectral dataset.

#### 3.8.2 NIR Spectra Pre-processing
*Wavelength Reduction:*
> The raw spectra were trimmed to the range 1050-1450 nm, where relevant chemical information to paracetamol was present.

*Standard Normal Variate (SNV)*
> Each spectrum ($X_{blend}$) was normalised to minimise scattering effects and baseline shifts. The transformation was applied as:

$$\mathbf{X}'_{blend} = \frac{\mathbf{X}_{blend} - \mathbf{X}_{blend,mean}}{\mathbf{X}_{blend,std.dev}}$$

*Savitzky-Golay Smoothing and Derivation*

A SG filter was applied to compute the first derivative of each spectrum, enhancing spectral features and reducing noise. Parameters used include:
- Window = 8
- Polynomial Order = 2
- Derivative Order = 1

### 3.8.3 Principal Component Analysis (PCA)

PCA was applied on the pre-processed spectra to reduce dimensionality and extract key patterns. The PCA transformation is described by:

$$\mathbf{T} = \mathbf{X}'_{blend} \cdot \mathbf{P}$$

where $\mathbf{T}$ represents the scores, $\mathbf{P}$ represents the loadings, and $\mathbf{X}'_{blend}$ is the pre-processed spectra matrix.

The first three principal components were retained, capturing 95% of the total variance in the spectral dataset.

### 3.8.4 Hotelling's $T^2$ Analysis

Hotelling's $T^2$ statistic was used to evaluate the multivariate distance of each spectrum from the PCA model centre. The $T^2$ value for each sample was calculated as:

$$T_i^2 = \sum_{j=1}^{k} \left( \frac{t_{ij}^2}{\lambda_j} \right)$$

where, $t_{ij}$ is the score of the *i*-th spectrum on the *j*-th PC, $\lambda_j$ is the variance of the *j*-th PC, and $k$ is the number of retained components (here, $k = 3$).

The control limit was determined based on chi-squared ($\chi^2$) distribution at a 99% confidence level:

$$Control\ Limit = \chi^2_{0.99,k}$$

For $k = 3$, the control limit was 11.34. Spectral outliers were identified where $T_i^2 > 11.34$. Detected outliers occurred at the following iterations: 3, 36, see Figure S15.

### 3.8.5 Integration with the Tableting DataFactory

NIR data collection and pre-processing were digitally integrated into the experimental workflow, enabling real-time spectral analysis. Spectra from 100 consecutive samples were recorded and processed, with visualisations generated for both raw and pre-processed data, see Figure S14. This allowed for the efficient detection of potential anomalies and enhanced confidence in the experimental outcomes.

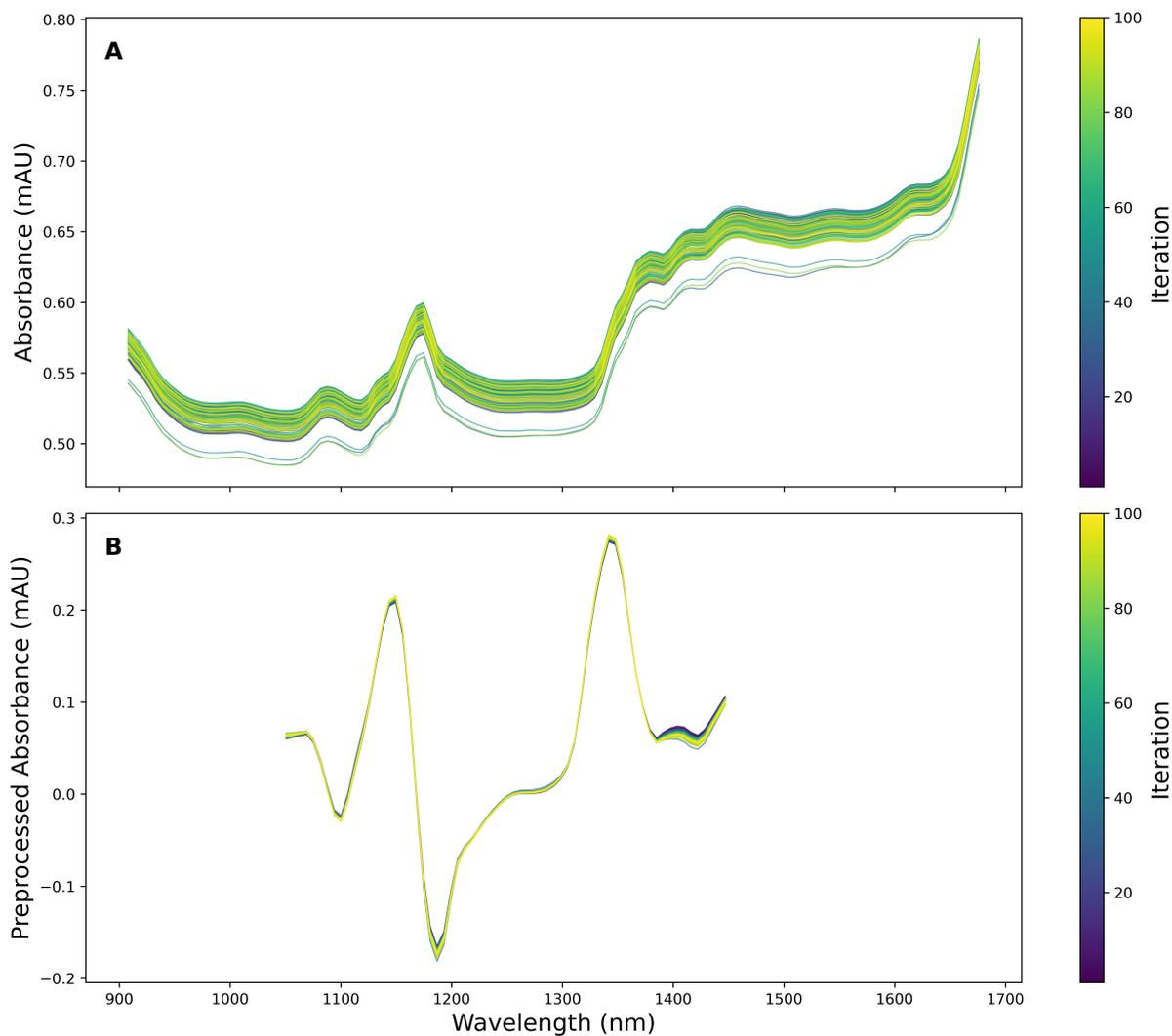

Figure S14: (A) Raw near-infrared (NIR) spectra collected from 100 consecutive samples, plotted as absorbance (mAU) across the wavelength range of 908–1676 nm. Each spectrum is colour-coded by its iteration number, with earlier samples shown in purple and later samples transitioning to yellow. (B) Pre-processed spectra of the same samples, following trimming to 1050–1450 nm wavelength range, standard normal variate (SNV) correction, and Savitzky-Golay smoothing and derivation. The colour coding corresponds to the same iteration numbers as in (A).

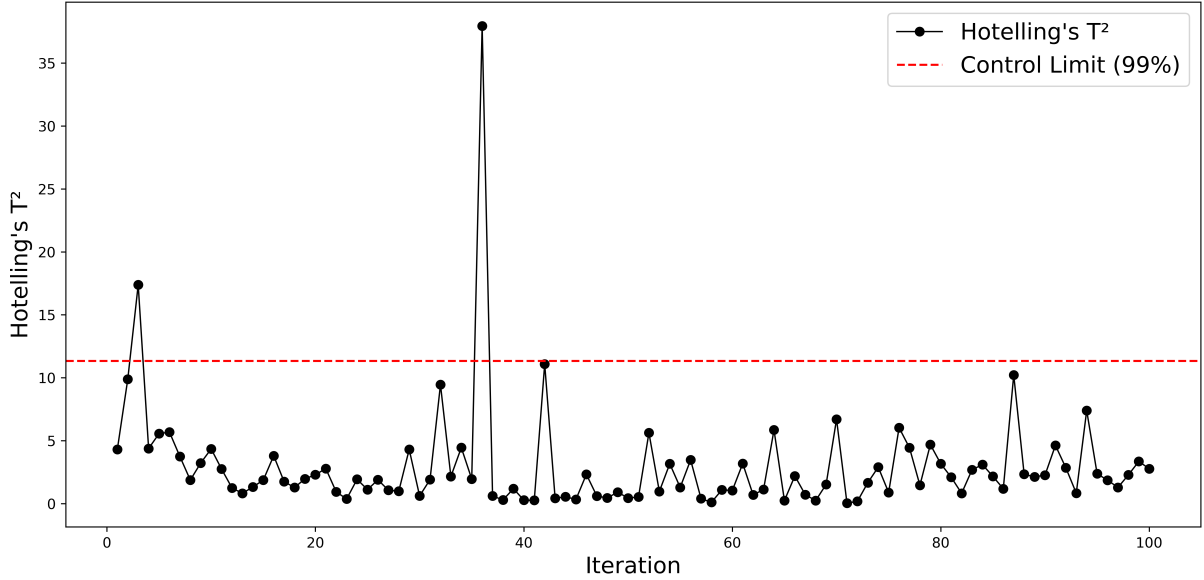

Figure S15: *Hotelling's $T^2$* chart for 100 consecutive samples, calculated using the first three principal components from the pre-processed NIR spectra. Each data point represents the $T^2$ statistic for an individual sample, quantifying its multivariate distance from the PCA model centre. The red dashed line indicates the control limit at the 99% confidence level ($T^2 > 11.34$).

## 4 Real-time Process Optimisers

### 4.1 Background of Bayesian Optimisation

Bayesian optimisation (BO) is a framework for optimising expensive-to-evaluate functions, particularly in high-dimensional or non-convex spaces where traditional optimisation methods may fail [1]. BO leverages a probabilistic surrogate model, often a Gaussian Process (GP), to model the objective function $\ell(\underline{X})$, where $\underline{X}$ is the state vector of input parameters. Given this prohibitively expensive objective function, the uncertainty of the objective $\ell(\cdot)$ across not-yet-evaluated input points is modelled as a probability distribution. BO models $\ell(\cdot)$ as a GP, which can be evaluated relatively cheaply and often with reasonable accuracy [2]. At each iteration the GP model is used to select the most promising candidate $\underline{X}^*$ for evaluation. The costly function $\ell$ is then only evaluated at $\ell(\underline{X}^*)$ in this iteration. Subsequently, the GP updates its posterior belief $\tilde{\ell}(\cdot)$ with the new data pair $(\underline{X}^*, \ell(\underline{X}^*))$, and that pair is added to the known experiment set $\mathcal{D}_n = \{(\underline{X}_i, \ell(\underline{X}_i))\}_{i=1}^{n}$. This iteration can be repeated to iterate to an optimum. The critical step is the selection of the candidate point $\underline{X}^*$, which is performed via an acquisition function that enables active learning of the objective $\ell(\cdot)$ [3]. The acquisition function $\alpha(\underline{X})$ guides the selection of the next evaluation point $\underline{X}_{n+1}$ by quantifying the utility of evaluating $\ell(\underline{X})$ at a given point, typically formulated as:

$$\underline{X}_{n+1} = \arg\max_{\underline{X}} \alpha(\underline{X}|\mathcal{D}_n)$$

This iterative process continues until convergence criteria are met. Common acquisition functions include the Expected Improvement (EI), Upper Confidence Bound (UCB), and Probability of Improvement (PI) [4]. In this study, EI is used the acquisition function due to its proven efficiency in balancing between the exploration and exploitation [5].

## 4.2 Digital integration

Both optimisation frameworks, PIBO and MOBO, were digitally integrated with the Tableting DataFactory through a local call mechanism between LabVIEW and Python scripts. LabVIEW triggered the Python scripts to execute the optimisation routines, ensuring seamless communication and data flow between the control system and the optimisation algorithms. Each experiment was repeated three times for enhanced consistency, with the average value from three tablets used in the optimisation workflow.

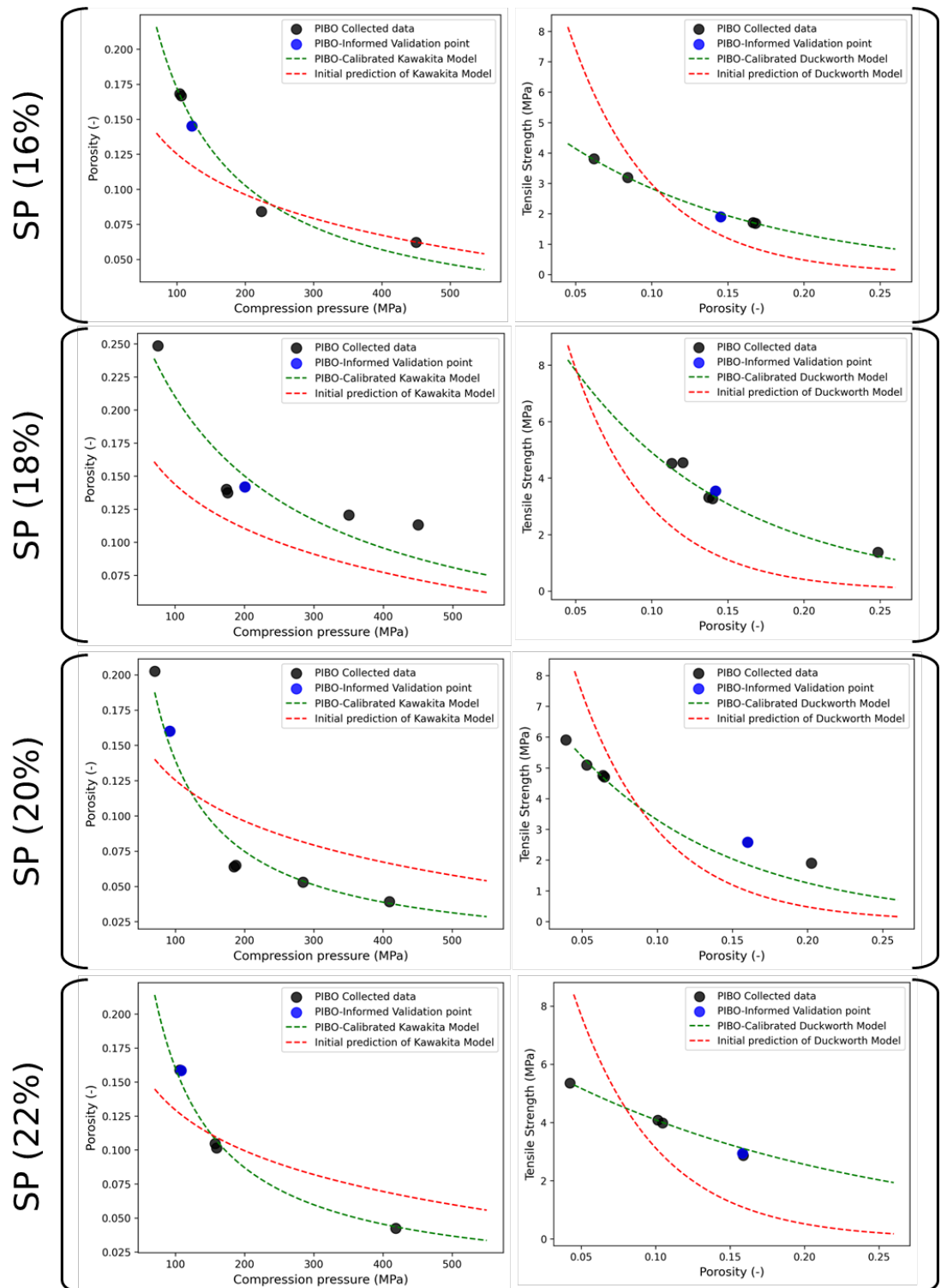

Figure S16: Initial and calibrated prediction of compressibility and compactability profiles

before (using the system of models) and after the calibration with PIBO.

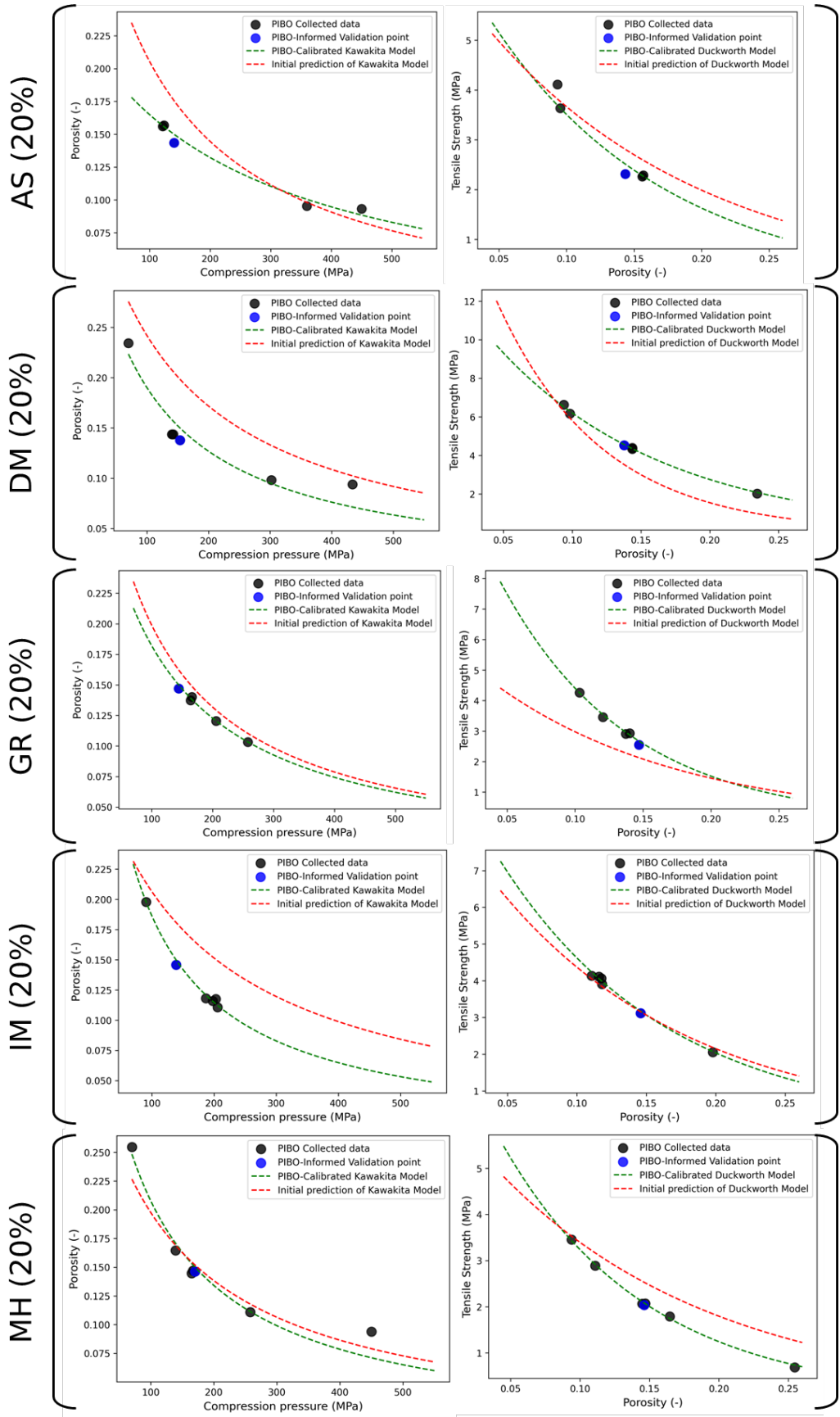

Figure S17 (Cont'd): Initial and calibrated prediction of compressibility and compactability profiles before (using the system of models) and after the calibration with PIBO.

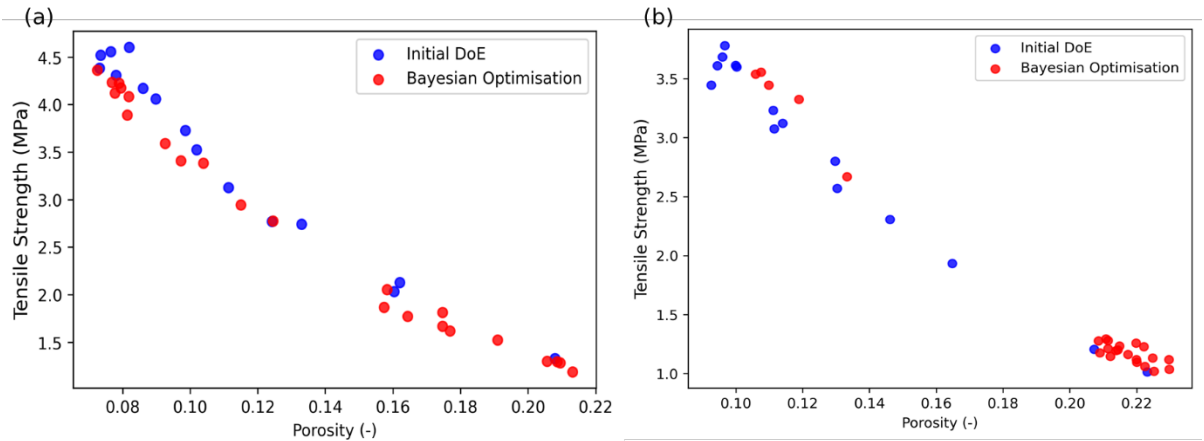

Figure S18: Representation of compactability profile of SP (20%) (a) and AS (20%) (b) formulations using the collected data points during MOBO.

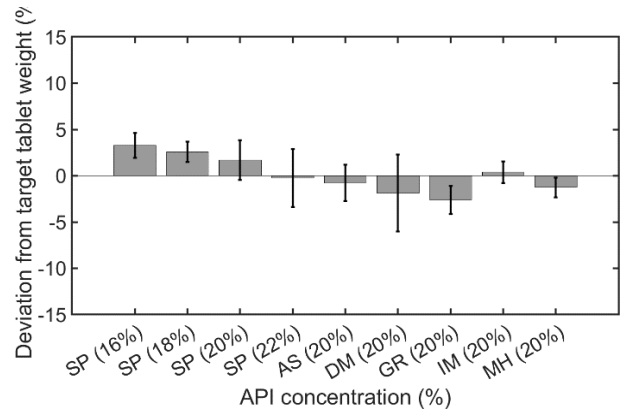

(a)

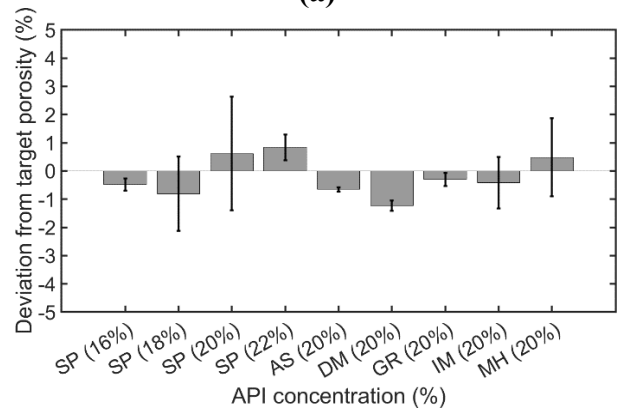

(b)

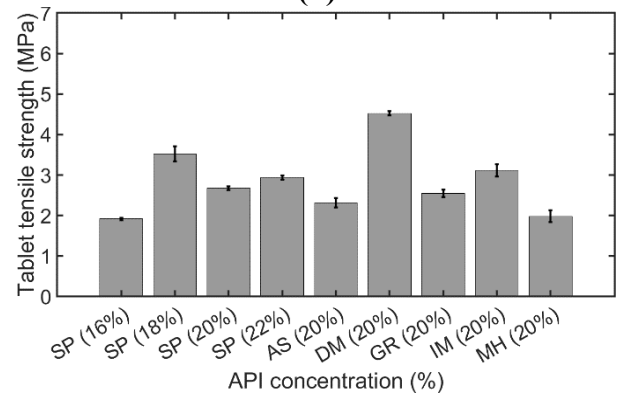

(c)

Figure S19: Validation of optimized target a) weight b) porosity c) tensile strength of tablets with different APIs and concentration.